%% file: MSarxiv.tex
\documentclass[12pt]{article}

\usepackage{times}

\usepackage{lineno}
	\nolinenumbers
\usepackage{graphicx}
\usepackage{nameref}
\makeatletter
\renewcommand\section{\@startsection{section}{1}{\z@}{0ex}{0ex}{\itshape}}
\makeatother

\begin{document} 
\newcommand{\gerda}       {\textsc{Gerda}}
\newcommand{\GERDA}       {\mbox{\textsc{Gerda}}}
\newcommand{\onbb}        {\ensuremath{0\nu\beta\beta}}
\newcommand{\qbb}         {\ensuremath{Q_{\beta\beta}}}
\newcommand{\gesix}       {{$^{76}$Ge}}
\newcommand{\gess}        {{$^{76}$Ge}}
\newcommand{\ctsper}      {cts/(keV$\cdot$kg$\cdot$yr)}
\newcommand{\kgyr}        {{kg$\cdot$yr}}
\newcommand{\nnbb}        {\ensuremath{2\nu\beta\beta}}
\newcommand{\thalfzero}   {${T^{0\nu}_{1/2}}$}
\newcommand{\fgesix}      {\mbox{$f_{76}$}}
\newcommand{\exposure}    {\mbox{$\cal E$}}
\newcommand{\majorana}    {\textsc{Majorana}}

\newenvironment{sciabstract}{%
\begin{quote} \bf}
{\end{quote}}

{\LARGE Probing Majorana neutrinos with  double-$\beta$ decay}
\\
\\
\centerline{(\gerda \ Collaboration)*}
\begin{center}
M.~Agostini$^{16}$,
A.M.~Bakalyarov$^{14}$,
M.~Balata$^{1}$,
I.~Barabanov$^{12}$,
L.~Baudis$^{20}$,
C.~Bauer$^{8}$,
E.~Bellotti$^{9,10}$,
S.~Belogurov$^{13,12}$,
A.~Bettini$^{17,18}$,
L.~Bezrukov$^{12}$
% ...
\end{center}
\normalsize{$^{1}$ INFN Laboratori Nazionali del Gran Sasso and Gran Sasso Science Institute, I-67100 Assergi, Italy.}\\
\normalsize{$^{2}$ INFN Laboratori Nazionali del Gran Sasso and Universit\`a degli Studi dell'Aquila, I-67100 L'Aquila, Italy.}\\
\normalsize{$^{3}$ INFN Laboratori Nazionali del Sud, I-95123 Catania, Italy.}\\
\normalsize{$^{4}$ Institute of Physics, Jagiellonian University, Cracow 40-348, Poland.} \\
\normalsize{$^{5}$ Institut f{\"u}r Kern- und Teilchenphysik, Technische Universit{\"a}t Dresden, D-01069 Dresden, Germany.} \\
\normalsize{$^{6}$ Joint Institute for Nuclear Research, Dubna 141980, Russia.} \\
\normalsize{$^{7}$ European Commission, JRC-Geel, B-2440 Geel, Belgium.} \\
\normalsize{$^{8}$ Max-Planck-Institut f{\"u}r Kernphysik, D-69117 Heidelberg, Germany.} \\
\normalsize{$^{9}$ Dipartimento di Fisica, Universit{\`a} Milano Bicocca, I-20126 Milano, Italy.} \\
\normalsize{$^{10}$ INFN Milano Bicocca, Milano, I-20126 Italy.} \\ 
\normalsize{$^{11}$ Dipartimento di Fisica, Universit{\`a} degli Studi di Milano e INFN Milano, I-20133 Milano, Italy.} \\ \newpage \noindent
\normalsize{$^{12}$ Institute for Nuclear Research of the Russian Academy of Sciences, Moscow 117312 , Russia.} \\ 
\normalsize{$^{13}$ Institute for Theoretical and Experimental Physics, Moscow 117259, Russia.} \\
\normalsize{$^{14}$ National Research Centre ``Kurchatov Institute'', Moscow 123182, Russia.} \\
\normalsize{$^{15}$ Max-Planck-Institut f{\"ur} Physik, D-80805 M{\"u}nchen, Germany.} \\
\normalsize{$^{16}$ Physik Department and Excellence Cluster Universe, Technische  Universit{\"a}t M{\"u}nchen, D-85748 Munich, Germany.} \\
\normalsize{$^{17}$ Dipartimento di Fisica e Astronomia dell{`}Universit{\`a} di Padova, I-35121 Padova, Italy.} \\
\normalsize{$^{18}$ INFN  Padova, I-35131 Padova, Italy.} \\
\normalsize{$^{19}$ Physikalisches Institut, Eberhard Karls Universit{\"a}t T{\"u}bingen, D-72076 T{\"u}bingen, Germany.} \\
\normalsize{$^{20}$ Physik Institut der Universit{\"a}t Z{\"u}rich, CH-8057 Z{\"u}rich, Switzerland.} \\
\normalsize{$\,*$ gerda-eb@mpi-hd.mpg.de}\\ $\ddagger$ Deceased. \\

\date{}

\baselineskip18pt 

\begin{sciabstract}
A discovery that neutrinos are not the usual Dirac but Majorana fermions, i.e. identical to their antiparticles, 
would be a manifestation of new physics with profound implications for particle physics and cosmology.
Majorana neutrinos would generate neutrinoless double-$\beta$ (\onbb) decay, a matter-creating process without the 
balancing emission of antimatter.  
So far, \onbb\ decay has eluded detection.  
The \gerda\ collaboration searches 
for the \onbb\ decay of \gesix\ by operating bare germanium detectors in an active 
liquid argon shield.  
With a total exposure of 82.4~kg$\cdot$yr, 
we observe no signal and derive a lower half-life limit of $T_{1/2} > 0.9 \cdot 10^{26}$~yr (90\% C.L.).  
Our $T_{1/2}$ sensitivity assuming no signal is $1.1 \cdot 10^{26}$~yr.
Combining the latter with those from other \onbb \ decay searches yields a sensitivity to 
the effective Majorana neutrino mass of 0.07 - 0.16~eV, with corresponding sensitivities to the
absolute mass scale in $\beta$ decay  of 0.15 - 0.44~eV, and to the cosmological relevant sum of neutrino 
masses of 0.46 - 1.3~eV.
\end{sciabstract}

\paragraph*{Introduction}
The experimental study of neutrinos commenced with their discovery by Cowan and Reines in 1956 \cite{Cowan:1992xc}, 
but only at the turn of the millennium experimental proof was established that the 3 known neutrino types (flavors) 
$\nu_\alpha$ ($\alpha$ = $e$,\,$\mu$,\,$\tau$) can convert from one kind to another \cite{Fukuda:1998mi,Ahmad:2002jz,kzen}.
This flavor oscillation is only possible if neutrinos have non-zero mass which is currently the only established
contradiction to the standard model (SM) of particle physics.
From tritium $\beta$ decay experiments \cite{Kraus:2004zw, Aseev:2011dq} 
and cosmological observations \cite{Tanabashi:2018oca} we know that their masses are very small - 
less than 10$^{-5}$ of the electron mass. 
Neutrinos are the only fundamental spin-1/2 particles (fermions) without electric charge. 
As a consequence, they might be identical to their antiparticles - named Majorana fermions{\footnote{Other fermions like the electron or
quarks are called Dirac particles.}} 
in honor of E. Majorana who suggested this first \cite{Majorana:1937vz}. 
This feature is a key ingredient for some explanations why matter is so much more abundant than antimatter in today's 
universe and why neutrinos are so much lighter than the other elementary particles \cite{moha2006}.

Majorana neutrinos would lead to nuclear decays which violate lepton number conservation and are therefore forbidden in the 
SM of particle physics. 
The so-called neutrinoless double-$\beta$ (\onbb) decay transforms  simultaneously two neutrons inside a nucleus into two protons with the 
emission of two electrons, see Fig.~1. 
The SM allowed double-$\beta$ ($2\nu\beta\beta$) decay occurs with the emission of two
electrons and two antineutrinos. 
In \onbb \ decay, the two electrons together carry the available decay energy (\qbb) 
and the resulting mono-energetic signal is the prime experimental signature. 
A positive detection would imply the first observation of a matter-creating process, without the balancing emission of antimatter,
and establish the Majorana nature of neutrinos \cite{SValle, Doro}.	

We report here on the search for 
the \onbb\ decay  $^{76}$Ge\,$\rightarrow$\,$^{76}$Se + 2e$^-$ ($Q_{\beta\beta} = 2039.061 \pm 0.007$~keV \cite{Mount:2010zz}) 
with the GERmanium Detector Array (\gerda). 
\gerda\ is the first experiment that surpasses a 
sensitivity{\footnote{The sensitivity is defined as the median limit expected from many repetitions of the experiment 
assuming no signal.}}  
for the \onbb\ decay half-life of $T_{1/2} \sim 10^{26}$~yr (90\% C.L.) and 
that operates in a background-free regime such that the expected number of background events is less 
than 1 in the energy region of interest at the final exposure \cite{nature544}.
This achievement together with the superior energy resolution of Ge detectors is crucial for the future
discovery of the signal. 
\input{FIGURE1.tex}

\paragraph*{Detector}
To achieve this goal, the \gerda\ experimental design was guided by the 
requirement to reduce interfering signals from naturally occurring radioactivity and from cosmic rays to negligible levels. 
The Ge material is  enriched in the \gesix\ isotope from the natural abundance of 7.8\%
to $>85$\% and transformed into high-purity Ge detectors. Thus the \onbb\ decay source and detector are 
identical as illustrated in Fig. 1. 
In total, \gerda \ deploys 37 enriched detectors with two different geometries (coaxial and Broad Energy Ge (BEGe) 
detectors) and with a total mass of 35.6~kg as bare crystals in 63~m$^3$ of liquid argon (LAr). 
The LAr serves as high-purity shielding against radiation from radioactive decays, and it also provides cooling for the Ge diodes. 
Moreover, the LAr  - due to its scintillation property - acts as a veto system to discard events originating from background radiation, 
which simultaneously deposit energy inside the Ge detectors and the adjacent LAr. 
The scintillation light is detected by 16 photomultipliers and wavelength shifting fibers connected to silicon photomultipliers.
A water tank encloses the LAr cryostat to further attenuate $\gamma$ radiation and neutrons from the experimental environment. 
It also serves as a water Cherenkov detector to identify cosmic ray muons and their secondary shower particles which otherwise 
could mimic signal events.
\gerda\ is operated deep underground at the Gran Sasso National Laboratories (LNGS) of INFN in Italy, at a depth of 3500 meter water 
equivalent to reduce the cosmic ray muon flux by six orders of magnitude with respect to the earth's surface. 
Detailed descriptions of Phases I and II of the experiment can be found in \cite{Agostini:2017hit} and in the 
supplementary materials.

\paragraph*{Performance}
The signals of the Ge detectors are read out by low-radioactive charge sensitive amplifiers, 
digitized at 100 MHz sampling rate and stored for off-line analysis. 
Weekly calibrations with $^{228}$Th sources are performed to monitor the energy scale and resolution, as well 
as to define and monitor the analysis cuts. 
The derived energy resolution, full width at half maximum (FWHM), at $Q_{\beta\beta}$ is $3.6\pm 0.1$~keV  for the coaxial and  
$3.0\pm 0.1$~keV  for the BEGe detectors, both corresponding to $\sigma / Q_{\beta \beta} < 10^{-3}$ ($\sigma$~=~FWHM\,/\,2.35).

During physics data taking all Ge and LAr scintillation channels are read out if one or more Ge 
diodes detect a signal above a preset trigger threshold. 
Multiple detector hits are discarded as background events. 
Similarly, events are classified as background if at least one photoelectron is detected in the LAr 
within $\sim 6~\mu$s around the Ge detector signal - that is 
$\sim 5$ times the lifetime of the argon excimer observed in \GERDA. 
Random coincidences lead to a loss of potential \onbb\  signals of $(2.3\pm 0.1)$\%. 
All events with a muon trigger preceding a Ge trigger by less than 10~$\mu$s are rejected with a signal 
loss of $<0.1$\%. 
Background events from $\gamma$ radiation lead often to multiple interactions within the same detector separated 
in space. The time structure of the recorded signal allows to reject this background as well as events occurring at 
the surface of a detector from $\alpha$ or $\beta$ decays (pulse shape discrimination, PSD).
More than 95\% of the background is rejected by the LAr veto and PSD (see Fig.~2) while 69\% 
of the \onbb \ decay events would be kept for the coaxial and 86\% for the BEGe detectors.
Compared to our previous publication \cite{Agostini:2018tnm}, 
the Phase~II exposure has been more than doubled while improving both energy resolution (by 10\%) and background rate 
(by $\sim$80\%) in the coaxial detectors and maintaining the excellent energy resolution of the BEGe detectors throughout
the run - altogether yielding a doubled sensitivity of more than 10$^{26}$~yr.

\paragraph*{Results}
Since its 
beginning \gerda\ has adopted a rigorous blind analysis strategy to ensure an unbiased search for \onbb\ decays. 
Events with a reconstructed energy of $Q_{\beta\beta} \pm 25$~keV are blinded, i.e. removed from the data stream, 
until the data selection is fixed.  
Fig.~2 displays the energy spectra corresponding to 53.9 kg$\cdot$yr Phase II exposure 
before and after analysis cuts, including a new PSD method for coaxial detectors (see supplementary material). 
\input{FIGURE2.tex}
At low energies the spectrum after analysis cuts is dominated by $2\nu\beta\beta$ decays. 
The insets in the figure display separately the event distribution of the coaxial and 
of the BEGe detector data sets in the analysis window 1930-2190\,keV. 
After unblinding, only 3 events in the coaxial and 4 events in the BEGe data sets remain in the analysis
window{\footnote{Three $\pm 5$~keV wide intervals at 2104 keV and 2119 keV, 
the position of known $\gamma$ lines, and  $Q_{\beta\beta}$ are excluded for the calculation of the background rate.
The limit calculation, however, includes the interval $Q_{\beta\beta} \pm 5$\,keV.}}.
\gerda\ thus reaches an unprecedented low background rate of 
$5.7^{+4.1} _{-2.6} \cdot 10^{-4}$~counts/(keV$\cdot$kg$\cdot$yr) 
for the coaxial  
and  
$5.6^{+3.4} _{-2.4} \cdot 10^{-4}$~counts/(keV$\cdot$kg$\cdot$yr) 
for the BEGe detectors.

An unbinned maximum likelihood fit is carried out simultaneously to the different data sets including those from \gerda\ Phase~I \cite{gerdaPRL2013}. 
In total 82.4~kg$\cdot$yr have been scrutinized for a \onbb\ signal so far. 
The fit function comprises flat distributions for the background, independent for each data set, and 
Gaussian distributions for a possible \onbb\ signal:
 the mean is $Q_{\beta\beta}$, the resolutions are
 taken from calibration data individually for each set, and 
 the normalizations are calculated from the searched for half-life $T_{1/2}$.
A null signal maximizes the likelihood. 
Confidence intervals are evaluated both in the frequentist and Bayesian frameworks. 
The frequentist analysis is based on the profile likelihood method and systematic uncertainties are included as 
nuisance parameters with Gaussian pull terms. 
The derived limit of $T_{1/2} > 0.9 \cdot 10^{26}$ yr (90\% C.L.) is compatible  with the  
sensitivity assuming no signal of $1.1 \cdot 10^{26}$~yr.  
\gerda\ is thus the first \onbb \ experiment to surmount  $10^{26}$ yr sensitivity.
The weaker limit is due to an event in the signal region at 2042.1~keV, 2.4 standard deviations ($\sigma$) away from \qbb .
The statistical analysis attributes it to background.
Statistical analysis including Bayesian inference is detailed in the supplementary materials. 

\paragraph*{Discussion}
Table~\ref{tab:limsens} compares our results with those of other \onbb \ decay searches. 
The $T_{1/2}$ sensitivities of other experiments are at most half of ours despite sometimes higher exposures.
\input{TABLE1.tex} 
This is due to \gerda 's lower background and superior energy resolution (see supplementary materials).
Several physical processes beyond the SM can produce \onbb \ decay. Here we focus on the paradigm of the mixing of
3 light Majorana neutrinos.
In this context, the half-life can be converted into a  \onbb \ decay strength that has the dimension of mass, denoted 
effective Majorana mass{\footnote{The unitary $3\times 3$ matrix 
    $U_{\alpha i}$ relates neutrino flavor states $\nu_\alpha$ ($\alpha$ = $e$,\,$\mu$,\,$\tau$) and mass eigenstates $\nu_i$~$(i=1,\,2,\,3)$.
    The absolute neutrino masses are still unknown, but two squared neutrino mass differences 
    $\Delta m_{21}^2$ and 
    $|\Delta m_{31}^2|$,\, ($\Delta m_{ij}^2 = \, m_i^2-m_j^2$),\,
    are known with increasing precision from neutrino oscillation experiments \cite{nufit2019}.}}\,
    $m_{\beta\beta} = |\sum\nolimits_{i=1}^3 U^2_{ei} m_i|$.
Nuclear structure details enter the decay rate, and uncertainties in the nuclear structure calculations result in a 
spread of $m_{\beta\beta}$ values for a given $T_{1/2}$ by typically a factor of 2-3 \cite{Engel:2016xgb}.
Some reported half-life limits $\cal L$ deviate by almost a factor of 2 from the associated sensitivity $\cal S$, 
indicating significant under-fluctuation (CUORE, KamLAND-Zen), or upward fluctuation (EXO-200). 
To overcome this possible behaviour of frequentist limits,
we use the sensitivity to extract the constraints on $m_{\beta\beta}$ shown in Table~\ref{tab:limsens}.
For \gerda \ the median limit is $m_{\beta\beta}< 0.1 - 0.23$~eV. 
Combining it with the sensitivities of the other searches (see supplementary materials) the bound tightens to 
$m_{\beta\beta}< 0.07 - 0.16$~eV (90\% C.L.), very similar to the bound deduced by KamLAND-Zen from their
$T_{1/2}$ limit \cite{kamlzen}.

Fig.~3 shows the dependence of the effective Majorana mass $m_{\beta\beta}$ as a function of the lightest neutrino 
mass $m_{light}$~=~min($m_i$), 
the cosmological observable of the sum of neutrino masses $\Sigma = {\sum\nolimits_{i} m_i}$, 
and the effective neutrino mass $m_{\beta} = \sqrt{\sum\nolimits_{i} |U^2_{ei}| m^2_i} $, i.e. the mass observable in single beta decays. 
\input{FIGURE3.tex}
The allowed parameter space is  classified according to the ordering of the neutrino mass eigenstates as 
normal ($\Delta m_{31}^2~>~0$) or inverted ($\Delta m_{31}^2~<~0$). 
The overlap region is called \lq quasi-degenerate\rq ; here the mass splittings are small compared to the absolute mass scale.
Latest oscillation data prefer normal ordering at the 3$\sigma$ level \cite{nufit2019}. 
Fig.~3 shows that our extracted limits of $m_{\beta\beta}$ disfavor a large fraction of the parameter space of quasi-degenerate 
Majorana neutrino masses. 
The combined limit of $m_{\beta\beta}$~=~0.16~eV corresponds to  constraints 
on $m_{light}<0.15-0.44$~eV, $\Sigma < 0.46-1.3$~eV and 
$m_{\beta} <  0.16-0.44$~eV.
Direct measurements of $m_{\beta}$ yield a limit of $\sim$2.3~eV \cite{Kraus:2004zw,Aseev:2011dq}.
In the upcoming years the KATRIN tritium decay experiment will increase the sensitivity to $\sim$0.2~eV \cite{katrin}. 
The sum of the neutrino masses influences the evolution and structure of the Universe. 
In the framework of the 6+1 parameter cosmological SM, the latest Planck data on the anistropy of 
the cosmic microwave radiation along with baryonic acoustic oscillation data provide limits as low as $\Sigma\,<\,0.12$~eV (95\%
C.L.) \cite{Aghanim:2018eyx}. Extended models relax these limits to $<$~0.37~eV ($<$~0.66~eV) for 1 (5) additional
parameters \cite{Tanabashi:2018oca}.	

Currently, there are no tensions among the 3 mass observables. A discovery of \onbb\ decay close to the current 
experimental half-life sensitivity should have counterpart signals in tritium $\beta$ decay and in cosmology, provided that 
the paradigm of 3 light Majorana neutrinos holds. In case of discrepancies with the other mass observables, 
a \onbb\  signal would point to other lepton number violating processes. 
Within the framework of 3 light Majorana neutrinos, the cosmological SM and in the absence of 
a \onbb\ decay at or close to the current sensitivity, the KATRIN experiment would not observe a signal. 
Conversely, a 
positive measurement of  $m_{\beta}>0.44$~eV in KATRIN would point to Dirac neutrinos or to an incomplete 
understanding of the nuclear physics \cite{Engel:2016xgb} of \onbb\ decay. 
It also would require extensions to the current minimal cosmological model. 
Instead, if the cosmological limit on $\Sigma$ holds,  \onbb\ decay experiments would have to probe a mass range 
$m_{\beta\beta} <0.05$~eV 
which requires a half-life sensitivity  of $10^{27}$~yr and above for a $^{76}$Ge-based experiment. 

The leading performance of  \gerda\  in terms of background suppression, energy resolution and sensitivity opens the way 
to {\sc Legend}, a next generation Ge experiment with sensitivity to half-lives of $10^{27}$~yr and beyond.  
A first phase 200~kg $^{76}$Ge experiment, 
{\sc Legend-200}  \cite{Abgrall:2017syy}, is under preparation at LNGS.

\vskip0.25truecm

\renewcommand{\refname}{\bf{References}}


\vskip0.25truecm

\noindent
{\bf ACKNOWLEDGEMENTS} \\ \noindent
The GERDA collaboration thanks the directors and the staff of the LNGS
for their continuous strong support of the GERDA experiment. 
{\bf Funding:} The GERDA
experiment is supported financially by the German Federal Ministry for Education and Research
(BMBF), the German Research Foundation (DFG) via the Excellence Cluster Universe and the
SFB1258, the Italian Istituto Nazionale di Fisica Nucleare (INFN), the Max Planck Society
(MPG), the Polish National Science Center (NCN) under grant UMO-2016/21/B/ST2/01094, the Foundation for Polish Science
(TEAM/2016-2/2017), the Russian Foundation for Basic Research (RFBR) and the Swiss
National Science Foundation (SNF). The institutions acknowledge also internal financial
support. This project has received funding or support from the European Union's Horizon 2020
research and innovation programme under the Marie Sklodowska-Curie Grant Agreements No.
690575 and No. 674896, respectively.
{\bf Author contributions:}
All authors contributed to the publication, being differently involved in the design and 
construction of the detector system, in its operation, and in the acquisition and analysis of data. 
All authors approved the final version of the manuscript. In line with collaboration policy, 
the authors are listed alphabetically.
{\bf Competing interests:} The authors declare no competing financial interests.
{\bf Data availability:} All data generated during this analysis and shown in this document
are available as png, pdf and root/txt files from the \gerda \ repository at Zenodo; 
http://doi.org/10.5281/zenodo.3339351. 
For further information contact the GERDA Collaboration (gerda-eb@mpi-hd.mpg.de).
\newline
\newline
\noindent



\clearpage
\newpage


\baselineskip18pt 

\title{Supplementary Materials for \\ Probing Majorana neutrinos with  double-$\beta$ decay}

\author{GERDA Collaboration}

\maketitle

\date{}

\renewcommand{\thefigure}{S\arabic{figure}}
\renewcommand{\thetable}{S\arabic{table}}
\renewcommand{\thesection}{S\arabic{section}}
\renewcommand{\thepage}{S\arabic{page}}
\setcounter{table}{0}
\setcounter{figure}{0}
\setcounter{section}{0}
\setcounter{page}{1}
\setcounter{linenumber}{1}


\paragraph*{Experimental setup}
 
The Germanium Detector Array (\gerda) experiment is located in Hall~A of the INFN Laboratori Nazionali del Gran Sasso; 
see ref.\cite{Ackermann:2012xja} for a detailed description. 
A rock overburden of about 3500\,m water equivalent eliminates the hadronic and electromagnetic components of the cosmic ray air showers and
reduces the cosmic muon flux to about 1.25/(m$^2\cdot$h).
The array of high-purity Ge detectors with $^{76}$Ge fraction enriched up to more than 85\% ($^{enr}$Ge) is operated 
in a 64\,m$^3$ cryostat filled with high-purity (N5.0) liquid argon~(LAr). 
The liquified noble gas serves both as coolant for the detectors as well as a shield against external background radiation.
The vacuum-insulated cryostat (\O\,4\,m) is produced from stainless steel of low radioactivity \cite{Maneschg:2008zz}. 
A 6~cm layer of low background copper on its inner wall shields the radioactivity of the steel. 
The shield  is complemented by 590\,m$^3$ of high-purity water (18\,M$\Omega\cdot$cm) contained in a tank (\O\,10\,m, h\,$=$\,9\,m) 
surrounding the cryostat.
This design has been shown to suppress the external $\gamma$ background at \qbb$=$2039\,keV to a level of
less than 10$^{-5}$\,\ctsper \cite{Barabanov:2009zz}.
Muons crossing the water tank are detected with 66 8" photo multiplier tubes (PMTs) by their Cherenkov radiation; 
together with plastic scintillator panels covering 4$\times$3\,m$^2$ on the roof of the experiment a muon rejection 
efficiency of $>$99.9\% is achieved \cite{Freund:2016fhz}. 

\begin{figure}[h]
	\centering
	\includegraphics[width=\textwidth]{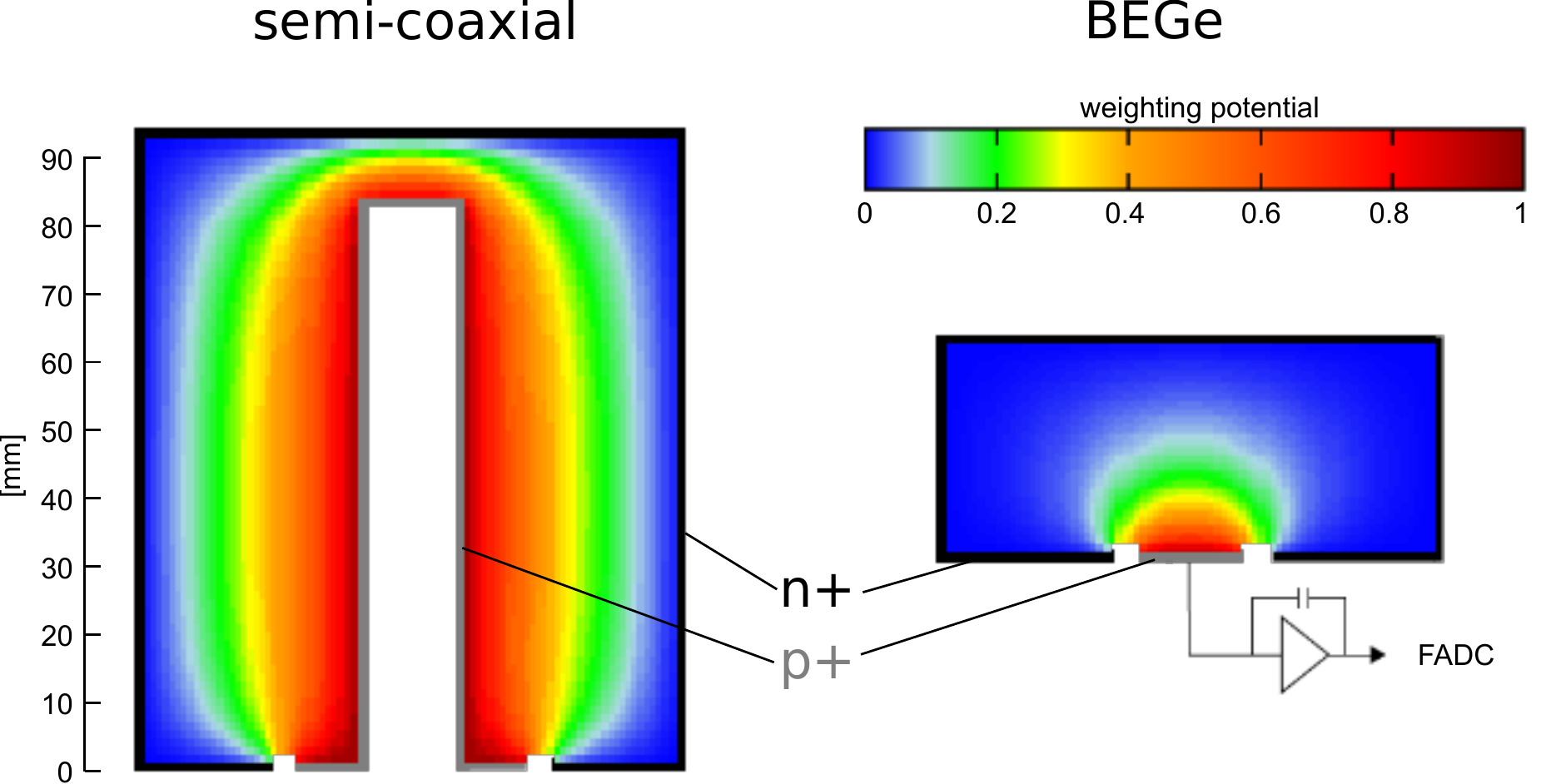}
\caption{
Cross sections of a \gerda\ coaxial and BEGe detector with an overlay of the corresponding weighting
potentials. Indicated are the n+ contact of diffused lithium (black)
and the p+ contact of ion-implanted boron (gray), separated by an insulating groove.
Overlayed in the active volume of the detectors is the weighting potential, which describes the coupling of the charge at a given location to the respective electrode.
For BEGe detectors, it is strongly peaked close to the small readout electrode, yielding to first order 
a pulse shape independent of the interaction position for most of the volume; the signal timing reflects the position-dependent drift time. 
This enables a simple discrimination between events with single
and multiple interactions in the detector. Such a discrimination is more complex in case of coaxial
detectors which feature a wider variety of signal shape patterns depending on the interaction position.
}
\label{fig:detlayout} 
\end{figure}

\gerda\ Phase\,I data taking started in November 2011 with a detector array consisting of 8 refurbished 
coaxial $^{enr}$Ge detectors from the Heidelberg-Moscow \cite{Gunther:1997ai} and {\sc Igex} \cite{Aalseth:2002rf} 
experiments (ANG 1-5, RG 1-3, 17.7\,kg total mass) and 3 coaxial high-purity Germanium (HPGe) detectors from low-background natural material 
(GTF). In July 2012, two GTF detectors were replaced by five Phase\,II type enriched Broad Energy Germanium (BEGe) detectors
(GD32B-D, GD35B-C, 3.6\,kg total mass), see Fig.~\ref{fig:detlayout} and Table~\ref{table:detproperties} for detector
layout and characteristics.

Phase\,I concluded in September 2013 after collecting an exposure of 23.5\,\kgyr\ without the observation of a 
\onbb\ decay signal, achieving a
background index at \qbb\ of $\sim$\,10$^{-2}$\,\ctsper\ and 
setting a 90\% C.L. limit on the half-life of T$^{0\nu}_{1/2}>2.1\times 10^{25}$\,yr (median sensitivity $2.4\times 10^{25}$\,yr)~\cite{gerdaPRL2013}. 

\begin{table}[htbp]
\begin{center}
\caption{ \label{table:detproperties}
Characteristics of the HPGe detectors (GDxyZ denoting the BEGe detectors) used in \gerda\ Phase\,II, for more details 
see \cite{Agostini:2017hit}. The table shows
the total and active mass (M$_{tot}$ and M$_{active}$), the enriched \gesix\ fraction (f$_{76}$), 
the detection efficiency  for the \onbb\ decay events ($\epsilon_{fep}$), as well as resolution FWHM and exposure 
$\mathcal{E}_{ana}$ used in analysis. 
The position number in a given detector string increases from top to bottom, String VII is in the central string of the 
array. 
The other strings are placed around the center following a hexagonal pattern.}
\begin{footnotesize}
\vspace*{2mm}
\begin{tabular}{rlccccccc}
\hline 
\hline \\[-2ex] 
Nr.	& Detector		& String-	& {M$_{tot}$}	& {M$_{active}$}& f$_{76}$	& $\epsilon_{fep}$ 	& FWHM 		& $\mathcal{E}_{ana}$ \\
    & 				& Position	& (g)			& (g)			&   	& 	& (keV) 	& (\kgyr)  \\
\hline
0	& GD91A			& I-0		& 627 			&	557(11) 	& 0.877(13)		& 0.898(2)			& 2.4(1)		& 1.1\\ 
1	& GD35B			& I-1		& 810 			&	740(12) 	& 0.877(13)		& 0.902(2)			& 2.6(1)		& 1.4\\ 
2	& GD02B			& I-2		& 625 			&	553(11) 	& 0.877(13)		& 0.895(2)			& 2.9(1)		& 1.2\\ 
3	& GD00B			& I-3		& 697 			&	613(13) 	& 0.877(13)		& 0.897(2)			& 3.0(1)		& 1.2\\ 
4	& GD61A			& I-4		& 731 			&	652(13) 	& 0.877(13)		& 0.902(2)			& 3.3(1)		& 1.3\\ 
5	& GD89B			& I-5		& 620 			&	533(13) 	& 0.877(13)		& 0.890(2)			& 3.8(1)		& 0.6\\ 
6	& GD02D			& I-6		& 662 			&	552(11) 	& 0.877(13)		& n.a.			& 4.4(1)		& 0.0\\ 
7	& GD91C			& I-7		& 627 			&	556(12) 	& 0.877(13)		& 0.896(2)			& 3.7(1)		& 0.2\\ 
\hline 
8	& ANG5			& II-0		& 2746 			&	2281(132) 	& 0.856(13)		& 0.918(18)			& 3.3(1)		& 5.0\\ 
9	& RG1			& II-1		& 2110 			&	1908(125) 	& 0.855(15)		& 0.915(18)			& 3.6(1)		& 3.8\\ 
10	& ANG3			& II-2		& 2391 			&	2070(136) 	& 0.883(26)		& 0.916(18)			& 3.4(1)		& 4.5\\ 
\hline 
11	& GD02A			& III-0		& 545 			&	488(9) 	& 0.877(13)		& 0.893(2)			& 2.4(1)		& 1.0\\ 
12	& GD32B			& III-1		& 716 			&	632(11) 	& 0.877(13)		& 0.900(2)			& 2.7(1)		& 1.2\\ 
13	& GD32A			& III-2		& 458 			&	404(11) 	& 0.877(13)		& 0.888(2)			& 3.3(1)		& 0.5\\ 
14	& GD32C			& III-3		& 743 			&	665(11) 	& 0.877(13)		& 0.901(2)			& 2.7(1)		& 1.4\\ 
15	& GD89C			& III-4		& 595 			&	520(13) 	& 0.877(13)		& 0.889(2)			& 3.1(1)		& 1.0\\ 
16	& GD61C			& III-5		& 634 			&	562(11) 	& 0.877(13)		& 0.892(2)			& 3.3(1)		& 1.0\\ 
17	& GD76B			& III-6		& 384 			&	326(8) 	& 0.877(13)		& 0.883(2)			& 3.3(1)		& 0.7\\ 
18	& GD00C			& III-7		& 815 			&	727(15) 	& 0.877(13)		& 0.903(2)			& 2.6(1)		& 1.4\\ 
\hline 
19	& GD35C			& IV-0		& 634 			&	572(10) 	& 0.877(13)		& 0.893(2)			& 2.3(1)		& 1.2\\ 
20	& GD76C			& IV-1		& 824 			&	723(13) 	& 0.877(13)		& 0.902(2)			& 2.6(1)		& 1.3\\ 
21	& GD89D			& IV-2		& 526 			&	454(10) 	& 0.877(13)		& 0.884(2)			& 2.7(1)		& 0.9\\ 
22	& GD00D			& IV-3		& 813 			&	723(14) 	& 0.877(13)		& 0.902(2)			& 2.8(1)		& 1.5\\ 
23	& GD79C			& IV-4		& 812 			&	713(12) 	& 0.877(13)		& 0.900(2)			& 3.6(1)		& 1.2\\ 
24	& GD35A			& IV-5		& 768 			&	693(13) 	& 0.877(13)		& 0.904(2)			& 3.1(1)		& 1.4\\ 
25	& GD91B			& IV-6		& 650 			&	578(11) 	& 0.877(13)		& 0.897(2)			& 4.1(1)		& 0.4\\ 
26	& GD61B			& IV-7		& 751 			&	666(13) 	& 0.877(13)		& 0.899(2)			& 3.0(1)		& 1.1\\ 
\hline 
27	& ANG2			& V-0		& 2833 			&	2468(145) 	& 0.866(25)		& 0.918(18)			& 4.1(1)		& 4.7\\ 
28	& RG2			& V-1		& 2166 			&	1800(115) 	& 0.855(15)		& 0.912(18)			& 3.8(1)		& 3.9\\ 
29	& ANG4			& V-2		& 2372 			&	2136(135) 	& 0.863(13)		& 0.916(18)			& 3.2(1)		& 4.4\\ 
\hline 
30	& GD00A			& VI-0		& 496 			&	439(9) 	& 0.877(13)		& 0.888(2)			& 3.0(1)		& 0.9\\ 
31	& GD02C			& VI-1		& 788 			&	700(14) 	& 0.877(13)		& 0.901(2)			& 2.7(1)		& 1.4\\ 
32	& GD79B			& VI-2		& 736 			&	648(14) 	& 0.877(13)		& 0.897(2)			& 2.9(1)		& 0.8\\ 
33	& GD91D			& VI-3		& 693 			&	615(13) 	& 0.877(13)		& 0.899(2)			& 2.8(1)		& 1.0\\ 
34	& GD32D			& VI-4		& 720 			&	657(11) 	& 0.877(13)		& 0.900(2)			& 3.0(1)		& 1.2\\ 
35	& GD89A			& VI-5		& 524 			&	462(10) 	& 0.877(13)		& 0.893(2)			& 3.2(1)		& 1.0\\ 
36	& ANG1			& VI-6		& 958 			&	795(50) 	& 0.859(29)		& 0.889(18)			& 3.4(1)		& 1.8\\ 
\hline 
37	& GTF112		& VII-0		& 2965 			&	2522(0) 	& 0.078(1)		& 0.920(18)			& 3.4(1)		& 0.0\\ 
38	& GTF32			& VII-1		& 2321 			&	2251(116) 	& 0.078(1)		& 0.920(18)			& 3.4(1)		& 0.0\\ 
39	& GTF45			& VII-2		& 2312 			&	1965(0) 	& 0.078(1)		& 0.920(18)			& 3.9(1)		& 0.0\\ 
\hline 
\hline 
\end{tabular}
\end{footnotesize}
\end{center}
\label{tab:detectors}
\end{table}

The current Phase\,II of the \gerda\ experiment started in December 2015. 
The many measures for reaching a sensitivity beyond T$^{0\nu}_{1/2}=10^{26}$\,yr at the design exposure of 100\,\kgyr \ 
are described in
detail elsewhere \cite{Agostini:2017hit}. The key is the further reduction of background events  from the array surroundings
in order to reach the \lq background-free\rq \ regime \cite{nature544}.
Major progress is due to (i) the deployment of additional $^{enr}$Ge detectors of the BEGe type (Canberra Semiconductor / Mirion) 
that exhibit superior pulse shape discrimination (PSD) and energy resolution \cite{Agostini:2013jta}, and (ii) the instrumentation of the LAr volume around the detector array with 
light sensors that allow for the active shielding (LAr veto) of events causing signals in both the Ge detectors and the LAr.      

\begin{figure}[htb]
	\centering
	\includegraphics[width=0.4\textwidth]{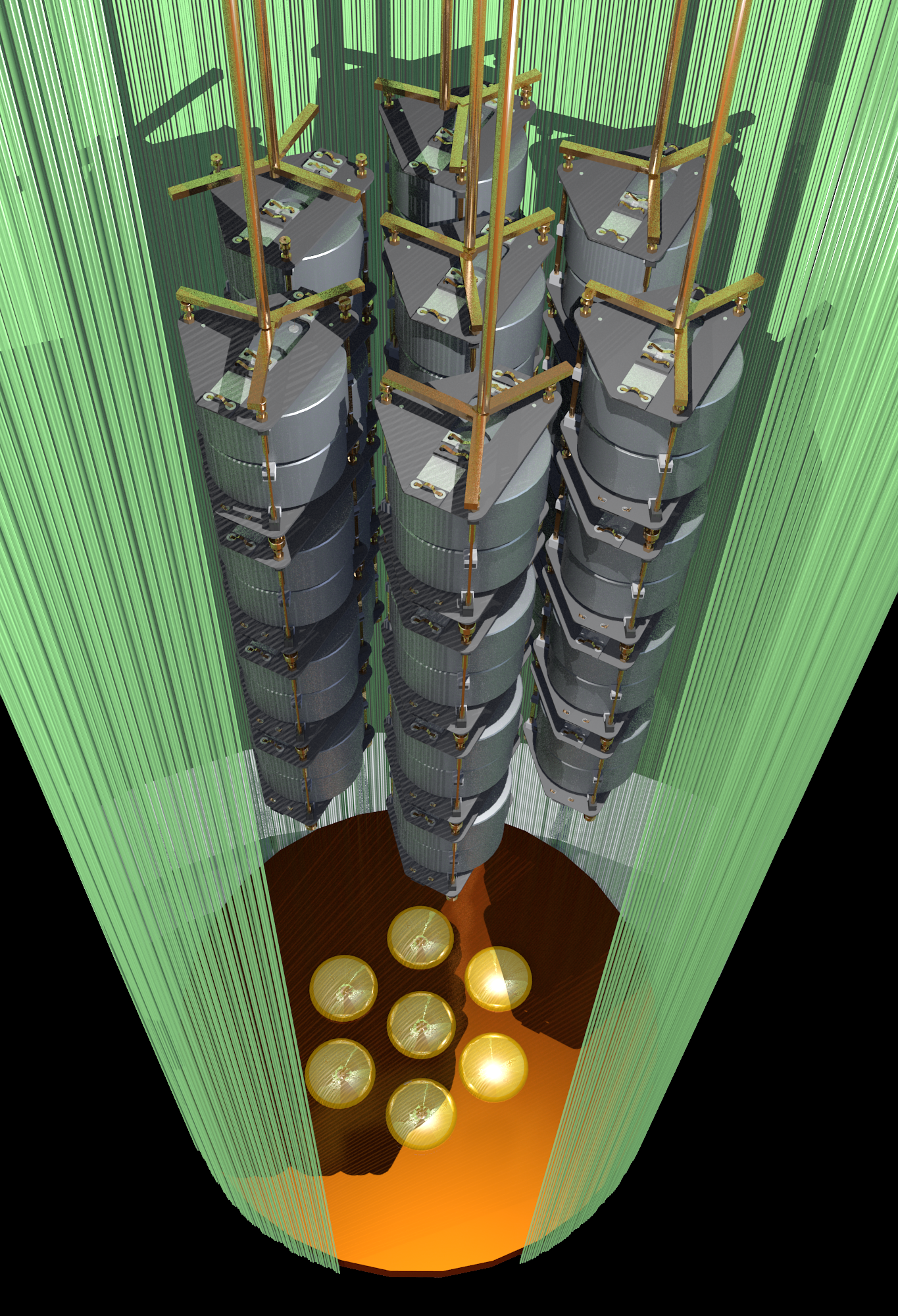}
	\caption{Schematic view of the Phase\,II germanium detector array and the liquid argon veto
system. The latter consists of a fiber curtain (green) read out on top by silicon
photo-multipliers (not shown), and 7 or 9 photo multiplier tubes at the bottom and top (not shown)
of the array, respectively. Omitted from the view are also the signal and high voltage cables of the individual
detectors as well as the transparent nylon shrouds that enclose each of the seven detector
strings in order to block the accumulation of $^{42}$K ions at the detectors.
	}
	\label{fig:setup} 
\end{figure}

Fig.~\ref{fig:setup} shows a schematic of the Phase\,II detector array and the surrounding LAr veto system.
The detector array is arranged in 7 strings consisting of 40 HPGe detectors: 
7 enriched coaxial detectors with a total mass of 15.6 kg, 3 coaxial detectors of natural isotopic abundance from Phase\,I, 
and in addition, 30 enriched BEGe detectors with a total mass of 20~kg (i.e. the GD detectors of Table~\ref{table:detproperties}).  
					
Both deployed detector types, coaxial and BEGe, are illustrated in Fig.~\ref{fig:detlayout}.
They both have cylindrical shapes and are fabricated from high purity Ge crystals with an active net impurity concentration 
around 
10$^{10}$ atoms\,/\,cm$^3$.
On the surface of the detectors is an n+ contact of diffused lithium and a p+ contact of ion-implanted boron, separated 
by a circular, non-conductive groove.
The n+ contact is 0.5 - 1.0\,mm thick and covers a major part of the detector surface.
It forms a \lq dead layer\rq, shielding the detector efficiently from all radiations that cannot penetrate this barrier,
like surface $\alpha$ particles, in particular.
The very thin p+ contact and the groove on the other hand do not provide such a shield.

A depletion region is formed by reverse biasing the p-n+ junction, i.e. by applying positive high
voltage to the n+ contact while keeping the p+ electrode grounded for signal read out.
Full depletion of the active region is typically reached at a few kilovolts.
Due to the electric field the electron-hole pairs created by ionizing radiation are separated and drift
into opposite directions. The moving charges induce mirror charges on the electrodes that are
read out by a charge sensitive preamplifier (as shown for the BEGe detector).

The various steps of the BEGe detector production from the procurement of the enriched material to the final working Ge diodes are described in  detail elsewhere \cite{Agostini:2014hra}. 
The average diameter and height of the detectors are 76\,mm and 30\,mm.
Further individual  characteristic parameters are collected in Table~\ref{table:detproperties}. 
The active masses and the detection efficiencies quoted in Table~\ref{table:detproperties} for
the new BEGe detectors are deduced from calibration measurements in vacuum at the HADES
facility \cite{Agostini:2014hra}.

The different layout of the readout electrodes of coaxial and BEGe detectors (Fig.~\ref{fig:detlayout}) is the source of their significantly different masses and energy resolutions. 
The bore-hole of the coaxial detector allows for the depletion of a large volume, i.e. for the
production of detectors with masses up to 3\,kg and beyond, while the small p+ contact restricts the mass of BEGe
detectors to less than 1\,kg. For larger masses a BEGe detector can not be fully depleted for voltages below
5~kV. On the other hand, the small readout electrode of a BEGe detector implies a
lower detector capacitance, 1\,pF vs. 30\,pF, and hence lower series noise resulting in superior energy
resolution. It is also the key for the superior PSD power of the BEGe detectors.

The signals of the Ge detectors are amplified
by low radioactivity, charge-sensitive preamplifiers \cite{Riboldi:2015dzj} of 30 MHz bandwidth, 
operated 35\,cm above the top of the detector array in the LAr.
The signals are led via $\sim$\,10\,m long coaxial cables to the outside of the lock where they are digitized (see next paragraph) 
\cite{Agostini:2017hit}.

The LAr veto system is a hybrid 
system evolved from studies of scintillation light detection in LAr with PMTs \cite{Agostini:2015boa} and silicon photo-multipliers (SiPMs) 
coupled to wavelength shifting fibers for increased light detection efficiency \cite{JanicskoCsathy:2010bh}.  Its central part (\O\,$=$\,0.5\,m, h\,$\approx$\,1.0\,m) 
consists of a curtain of 1$\times$1\,mm$^2$ multiclad fibers (Saint Gobain, BCF-91A) which are vacuum-coated with wavelength shifting
tetraphenyl butadiene (TPB). 
The fibers are arranged such that their geometrical coverage is 78\% by design.
Observed misalignments indicate, however, that coverage might be as low as 50\%.    
In total 405 fibers of about 1.8\,m length are deployed. Groups of nine fibers are bent in a U-shape such that both ends can be coupled on top 
of the curtain to two 3$\times$3\,mm$^2$ SiPM arrays (Ketek).
For readout, six SiPMs are connected in parallel to a 50\,$\Omega$ coaxial cable which transmits the signals to the outside for amplification
in a charge sensitive preamplifier. 
Additionally, 16 low-radioactivity PMTs (Hamamatsu, 3" R11065-10/20 MOD) with a TPB coated entrance window are installed in the cryostat.
Groups of 9 and 7 PMTs are located in a distance of about 1\,m from the Ge detectors above and below the array, respectively.
The PMTs and the fiber/SiPM assembly are mounted in a low-mass copper frame which can be deployed 
together with the detector array into the cryostat.

In Phase\,I, each detector string was enclosed by a copper cylinder of 103\,mm diameter in order to minimize the volume 
from which ions of the $\beta^-$ emitter $^{42}$K ($T_{1/2}=12.3$\,hr, 
$Q$-value\,$=$\,3.5\,MeV), progenies of $^{42}$Ar decays, can be collected at the detector surfaces.
For Phase\,II, these shrouds have been replaced by transparent ones made from nylon
coated with TPB \cite{Lubashevskiy:2017lmf}. This allows for the detection of scintillation light also from inside the volume enclosed by the shroud.


\paragraph*{Data processing and quality}

The data stream from the \gerda\ experiment comprises of data from the Ge-detector array, 
the LAr-veto system and the muon-veto system, see Tab.~\ref{tab:ADC} 
for a summary of detector systems and their data acquisition.
The signals of the Ge detectors are digitized individually by a 100\,MHz flash analog-to-digital converter (FADC). 
When a signal in any Ge channel crosses the trigger threshold, a main
trigger is issued and all Ge channels are read out. 
Additionally, a software trigger is issued about every 40\,s, irrespective of the signal in the Ge detectors,  
and test pulses are injected into all preamplifiers simultaneously every 20\,s.
For each event and each Ge channel a 160\,$\mu$s trace, downsampled to 25\,MHz and centered on the trigger time, is written on disk. 
The traces typically start with 80\,$\mu$s of baseline, followed by a sharp
leading edge of less than 1\,$\mu$s induced by the collection of the
electron-hole pairs generated in the detector during an event, and end with a $\sim$80\,$\mu$s exponential tail due to the decay time constant of the preamplifier.
For PSD analysis a 10~$\mu$s wide window around the leading edge is recorded at 100~MHz.

The data read out of the LAr veto devices is also initialized by the main trigger.
The signal of each PMT above/below the array
is digitized at 100\,MHz as an independent channel (12\,$\mu$s long traces). 
For the read out of the fiber curtain surrounding the Ge-detector array, 
9 fibers are coupled to one SiPM at each end, and groups of 6 SiPMs are summed in one amplifier and FADC channel.
For each trigger a 120\,$\mu$s trace is digitized with a sampling frequency of 12.5\,MHz.

The data acquisition of the muon veto operates independently of the main trigger from the Ge-detector array.
A trigger is issued if 5 PMTs in the water tank are above a threshold of 0.5\,p.e. or 
if there is a triple coincidence in the 3 layers of plastic scintillators on the roof above the experiment.
The status of a logic signal triggered by the muon veto is recorded with each main trigger in the data stream of Ge detectors and used as muon veto flag.

\begin{table}[tbp]
\begin{center}
\caption{ \label{tab:ADC}
Summary of GERDA Phase II detector systems with sensor types, number of sensors $N_{\mathrm{sens}}$, number of channels $N_{\mathrm{ch}}$, saved trace length and sampling frequency.
The signals are digitized by two data acquisition systems (14x8 channels) based on the SIS3301 FADC (100\,MHz, 14 bit depth), 
one for the muon veto systems and one for the remaining systems.
}
\vspace*{5mm}
\begin{tabular}{llccc}

\hline 
\hline \\[-2ex] 
Detector system			& Sensor			& $N_{\mathrm{sens}}$/$N_{\mathrm{ch}}$	& trace length 			& sampling frequency \\ 
						&					&										& ($\mu$s)				& (MHz)						\\
\hline
Ge detector array		& Ge diode 			& 40~/~40 		&	160 		& 25					\\	
		                &  			        &  		        &	10 		& 100					\\	
LAr veto PMTs			& 3''\,PMT 			& 16~/~16 		&	12 		& 100					\\ 
LAr veto fibers			& 3x3\,mm$^2$ SiPM 		& 90~/~15 		&	120 		& 12.5					\\ 
\hline
Muon veto water			& 8''\,PMT 			& 66~/~66 		&	4 		& 100					\\ 
Muon veto plastic		& 8\,mm PMT 			& 40~/~40 		&	4 		& 100				\\ 
\hline 
\hline 
\end{tabular}
\end{center}
\end{table}

All FADC data are converted in a ROOT-based data format for the
analysis~\cite{Agostini:2011nf}. The digital signal processing of the traces is
performed within the GELATIO software framework~\cite{Agostini:2011xe}.
Individual traces are analyzed to extract the parameters of 
interest for the event reconstruction~\cite{Agostini:2011mh}. Of particular 
interest is the amplitude of the Ge
pulses, which is proportional to the energy released in the detector. The
amplitude is reconstructed using a run-by-run optimized Zero Area Cusp (ZAC)
filter~\cite{Agostini:2015pta}. The calibration of this energy estimator is described in detail in the next section.


To remove non-physical events due
to electric discharges or a burst of noise, a suite of cuts is applied to the
parameters extracted during the digital signal processing.
For example the baseline, the leading edge and the decay tail of the pulse are required to satisfy a set of
heuristic conditions. If a trace does not pass the data cleaning, the full
event is tagged as non-physical. The probability of rejecting physical events has
been estimated to be below 0.1\% based on test-pulses
injected in the data stream. At the same time, the probability of accepting
non-physical events is expected to be below 0.1\% according to a visual
inspection of all events in the data set with energy between 1.6 and 3\,MeV.

As a last step of the data reduction the PSD observables are
calibrated and the selection criteria are evaluated.
The events are associated to different data sets and for each of them the live time and
signal-detection efficiencies are computed as exposure weighted average.

The \gerda\ collaboration enforces a strict blind analysis policy.
Events containing a Ge signal of energy \qbb$\pm$25\,keV are
automatically removed from the data (analysis) stream. Once all analysis cuts and
procedures are fixed, these events are processed in a user-protected area and
the traces are inspected. 

\begin{figure}[tb]
   \centering{
   \includegraphics[width=\textwidth]{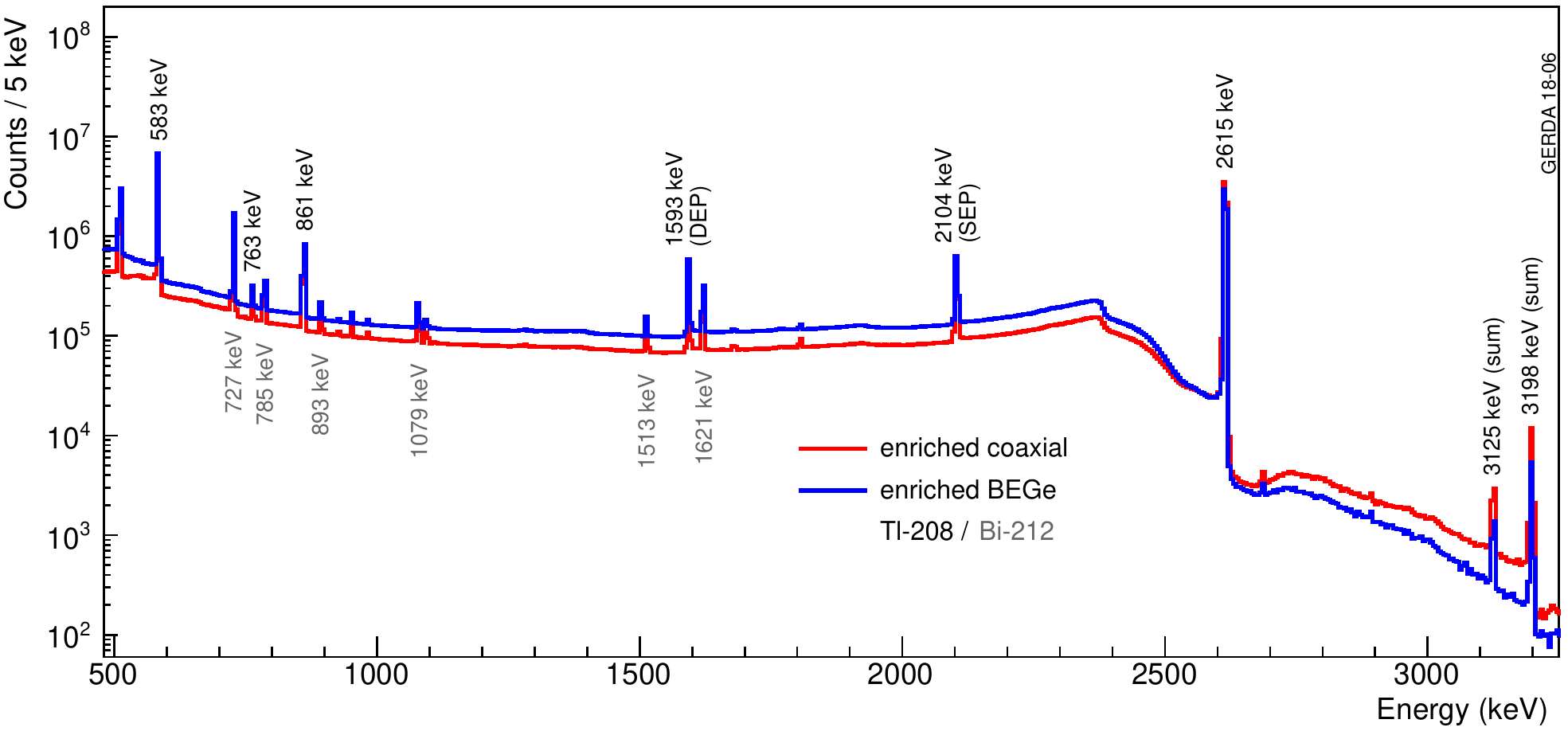}
}
   \caption{  
      The summed energy spectra of the weekly $^{228}$Th calibration runs for the coaxial (red) and BEGe (blue) detectors.
   }\label{fig:calib}
\end{figure}

\paragraph*{Calibration}
For calibrating the energy scale and determining the resolution of each germanium detector in \gerda, 
weekly calibrations are performed by lowering three $^{228}$Th sources of low neutron emission
with an activity on the order of 10\,kBq into the cryostat \cite{Baudis:2013kaa,Baudis:2015sba}. The sources are placed around the
detector array and moved to 3 positions along the array height to ensure a homogeneous
irradiation of all detectors. During a calibration run, a few thousand events are
accumulated in the 2615\,keV $\gamma$ line for each detector.
The complete set of calibration data collected during \gerda\ Phase\,II (about one
hundred calibration runs) is shown in Fig.~\ref{fig:calib} for the BEGe and
coaxial detectors separately.

The calibration of the energy scale is performed on the recorded energy spectrum for each detector and calibration run independently.
First, each calibration peak is fitted to identify its centroid, and then a linear
calibration function is extracted comparing the peaks centroids with the
expected values from literature.
The function used to fit the peaks is composed of a Gaussian
with centroid $\mu$ and resolution $\sigma$
superimposed to different functions to model the background (flat, linear, step).
For strong $\gamma$ lines, for example the 583\,keV and 2615\,keV lines, a third term
$h(E)$ is added to the fit model, with additional free parameters $c$ and $\beta$, to account for the low energy tail due to
ballistic deficit:
\begin{equation}
h(E) = \frac{c}{2\beta}\exp\left(\frac{E-\mu}{\beta}+\frac{\sigma^2}{2\beta^2}\right)\mathrm{erfc}\left(\frac{E-\mu}{\sqrt{2}\sigma}+\frac{\sigma}{\sqrt{2}\beta}\right)\mathrm{.}
\label{eq:tail}
\end{equation}
Typically, about 5-8 peaks between 583\,keV and 2615\,keV can be accurately fitted for each detector and
calibration run.
The extracted calibration curves are applied to each detector individually and for the
time following a calibration run until the next one. 

The stability of the energy scale is monitored using the test pulses injected every 20\,s in the
detector preamplifiers and the position of the 2615\,keV line between successive calibrations.
During stable periods, shifts in the 2615\,keV line are typically smaller than 1\,keV.
If instabilities on the order of 1\,keV or above are observed, the corresponding data is excluded from analysis, and a calibration run is performed immediately. 
This strict approach is needed to preserve the excellent energy resolution of
the Ge detectors when the data collected over an extended period of time 
are combined into a single data set.

\begin{figure}[tb]
\centering
	\includegraphics[width=\textwidth]{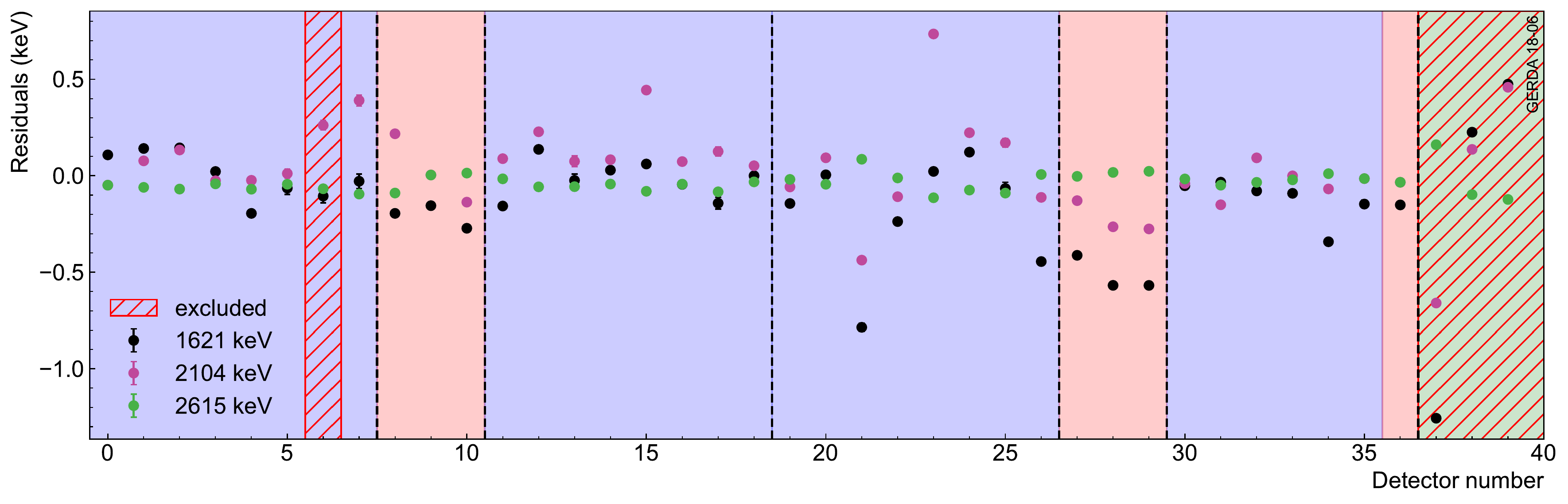}	
	\caption{
	The difference of reconstructed $\gamma$ line energies relative to literature 
	from the combined calibration spectra of the various detector types in \gerda, 
	BEGe (highlighted blue), coaxial (highlighted red) type and natural (highlighted hatched green,
 not considered for $0\nu\beta\beta$ decay analysis) detectors. Detector no.~6 (GD02D) has been excluded from analysis due
 to incomplete depletion.
}
	\label{fig:residuals}
\end{figure}

The different assumptions and approximations used in the calibration procedure can result in minor biases of the energy scale at $Q_{\beta\beta}$. 
For example, a linear calibration function describes the energy scale only approximately due to the non-linearities in the read-out chain.
The overall biases can be estimated by analyzing the peak positions in the combined calibration spectrum.
The combined calibration spectrum is obtained by summing up calibrated spectra of all calibration runs for each detector separately.
Fig.~\ref{fig:residuals} shows the residuals of prominent peaks in the combined spectrum for the different detectors. 
Close to the 583\,keV and 2615\,keV peaks the residuals are typically around 0.05\,keV, as these two are the strongest lines that have the largest weight in the fit of the calibration function.
Between these two lines the residuals are larger, typically around a few tenths of keV. 
We estimate a systematic uncertainty of 0.2\,keV on the energy scale at $Q_{\beta\beta}=2039$ for both the BEGe and coaxial
datasets using the average absolute deviation of individual detectors from the datasets at the closest spectral line at
2104\,keV.
The impact of this uncertainty on the derived limit for \onbb\ decay is discussed in the section 
on the statistical analysis.

When the data from different detectors are combined into a single data set, 
a peak should be modeled as a sum of multiple Gaussian distributions with different centroids and resolutions 
(i.e. a Gaussian mixture). In our case the resolution of the detectors in a data set varies by less than a factor of 2 and 
the shift between the centroids is much smaller than the energy resolution (biases are around 0.2\,keV). 
Therefore the shape of a peak in our data sets is approximately Gaussian and can be characterized by an effective resolution $\sigma$
that can be computed from the resolution of the individual detectors $\sigma_i$. 

Energy resolutions of individual detectors $\sigma_i$ are obtained from the best fit of a peak model to the peaks of the respective combined calibration spectrum, 
using the same method as described earlier in this section for individual calibration runs.

\begin{figure}[tb]
\centering
	\includegraphics[width=0.7\textwidth]{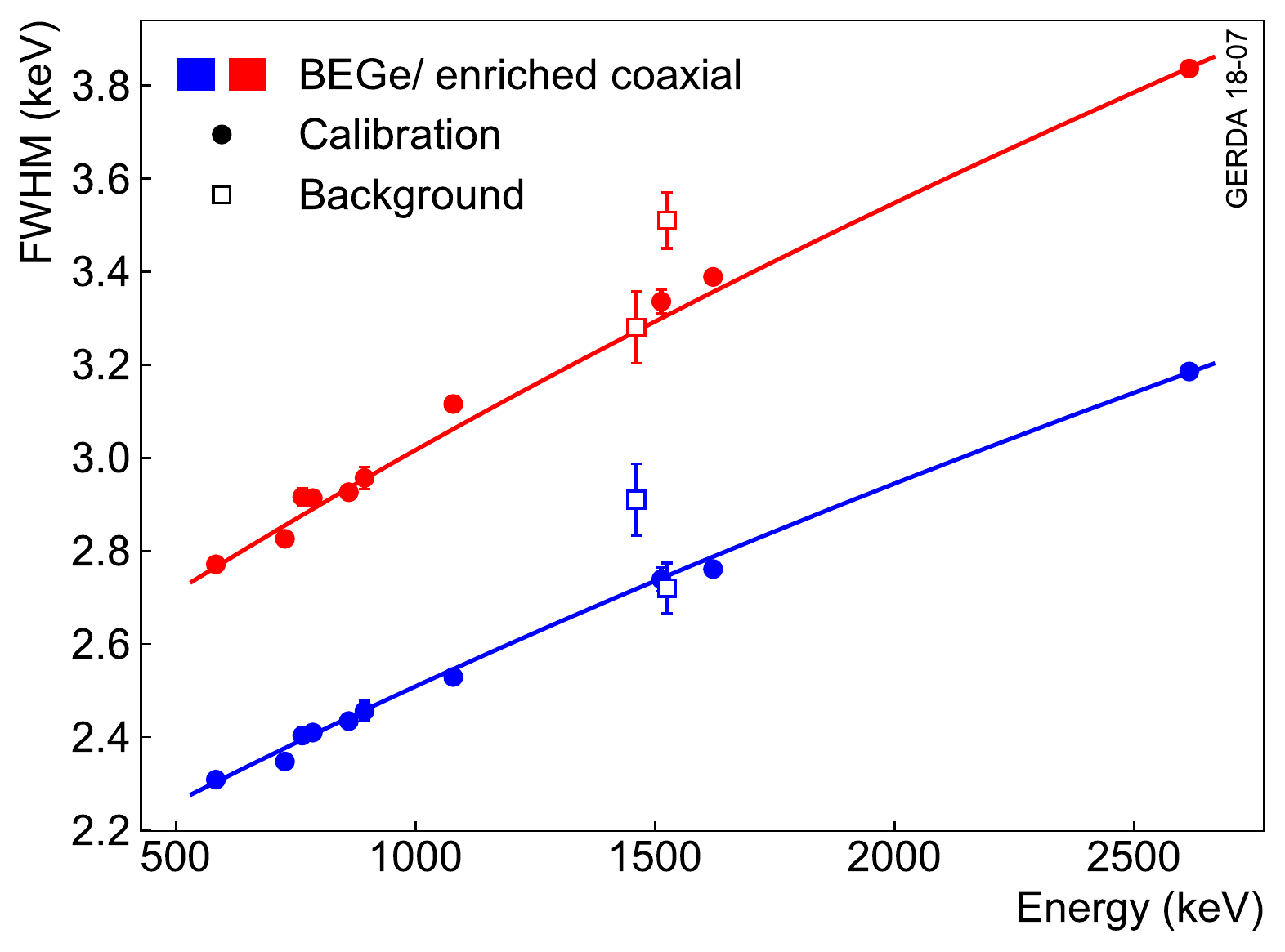}
	\caption{The effective Phase II energy resolution for all BEGe and coaxial detectors as a function of energy.
	Resolutions for each point are first determined from the summed spectrum of all calibrations for the individual detectors and then combined according to their respective exposure
	contribution in the final physics data set.
	The curves exhibit a systematic uncertainty of 0.1\,keV. 
	For comparison also the average resolutions of the strongest $\gamma$ lines in physics data, from $^{40}$K and $^{42}$K, 
	determined for individual detectors and then combined according to their exposure, are shown.
}
	\label{fig:resolutions}
\end{figure}

For each data set, the effective resolution $\sigma$ for each $\gamma$ line is then obtained with
\begin{equation}\label{eq:simpgaussmix}
   \sigma = \sqrt{\frac{1}{\mathcal{E}}\sum_i \mathcal{E}_i \cdot \sigma_i^2},
\end{equation}
where the index $i$ runs over the individual detectors of that data set, with the respective resolutions $\sigma_i$, exposures $\mathcal{E}_i$, and the total exposure $\mathcal{E}$. 
The resolution for the data sets of the two detector types deployed in \GERDA, BEGe and coaxial, are shown in Fig.~\ref{fig:resolutions} in terms of full width at half maximum ($\mathrm{\textrm{FWHM}}=2.35\cdot\sigma$).

To interpolate the resolution of the data set at $Q_{\beta\beta}$, the effective resolution of the $\gamma$ lines is fitted with the function: 
\begin{equation}
\label{eq:resfkt} 
\sigma(E) = \sqrt{a + bE}.  
\end{equation}
This function is a common choice for Ge-detectors and combines contributions to the resolution from electronics noise and fluctuations in the number of produced charge carriers.
The best fits are shown in Fig.~\ref{fig:resolutions}. 
Excluded from the fit are the single and double escape peaks of the 2615\,keV line of $^{208}$Tl, at 2104\,keV and 1593\,keV respectively, 
the annihilation peak at 511\,keV, and the summation peaks beyond 2615\,keV. 
These peaks feature additional effects that broaden them and make them not representative.

The estimated effective resolution at $Q_{\beta\beta}$ in terms of FWHM is 3.0(1)\,keV for the BEGe data set and 
3.6(1)\,keV for the coaxial data set.
The uncertainty in the resolution is dominated by contributions from the stability of the energy scale in between calibration runs.

The same approach can be used to interpolate the resolution of single detectors in their respective combined calibration spectra.
The FWHM of the individual detectors at $Q_{\beta\beta}$ is listed in Table~\ref{table:detproperties}; 
for BEGe detectors it is in the range 2.3-4.1\,keV, with a standard deviation of 0.4\,keV, and for the coaxial 
detectors, the range is 3.2-4.1\,keV, with a standard deviation of 0.3\,keV.

%
As a crosscheck, the resolution of $\gamma$ lines in the physics data was compared to the calibration data.
The resolutions of the most prominent lines, the $^{40}$K line at 1460\,keV and $^{42}$K line at 1525\,keV, are also shown in Fig.~\ref{fig:resolutions}.
To account for inhomogeneities in the count rate,
these resolutions are obtained by determining the resolution for each detector individually and 
then combining them by their exposures.

For both data sets, the resolution of one of the lines matches the expectation, while the other deviates moderately from the calibration data.
In previous \gerda\ analyses, the resolution of the coaxial detectors at \qbb\ was increased to match 
the resolution of the $^{42}$K line \cite{nature544}. 
 However, it was found that the worse resolution originates mainly from a strong mismatch of count rates in the 
 lines relative to individual exposures.
 Taking this into account, the remaining $\sim$0.2\,keV deviations are compatible with the systematic uncertainties on the
 resolution of 0.1\,keV.
 

\paragraph*{Liquid argon veto}

The read-out of the LAr scintillation light provides a tool to identify background events in which energy is simultaneously deposited 
in the Ge detectors and in the LAr.
The PMT and SiPM waveforms are processed offline to reconstruct the timing and amplitude of the scintillation signals.
An event is vetoed if a scintillation signal with amplitude above threshold is found in a narrow time window around the Ge pulse.
The threshold and time window is optimized channel-by-channel, taking into account the noise level and dark rate as well as the temporal structure of the LAr scintillation process.
The thresholds are between 0.2 and 0.9~photoelectrons, while the time windows have a width of about 6~$\mu$s and range 
between -0.8 to 6.0~$\mu$s relative to the Ge pulse.

\begin{figure}[tb]
    \centering
	\includegraphics[width=0.66\textwidth]{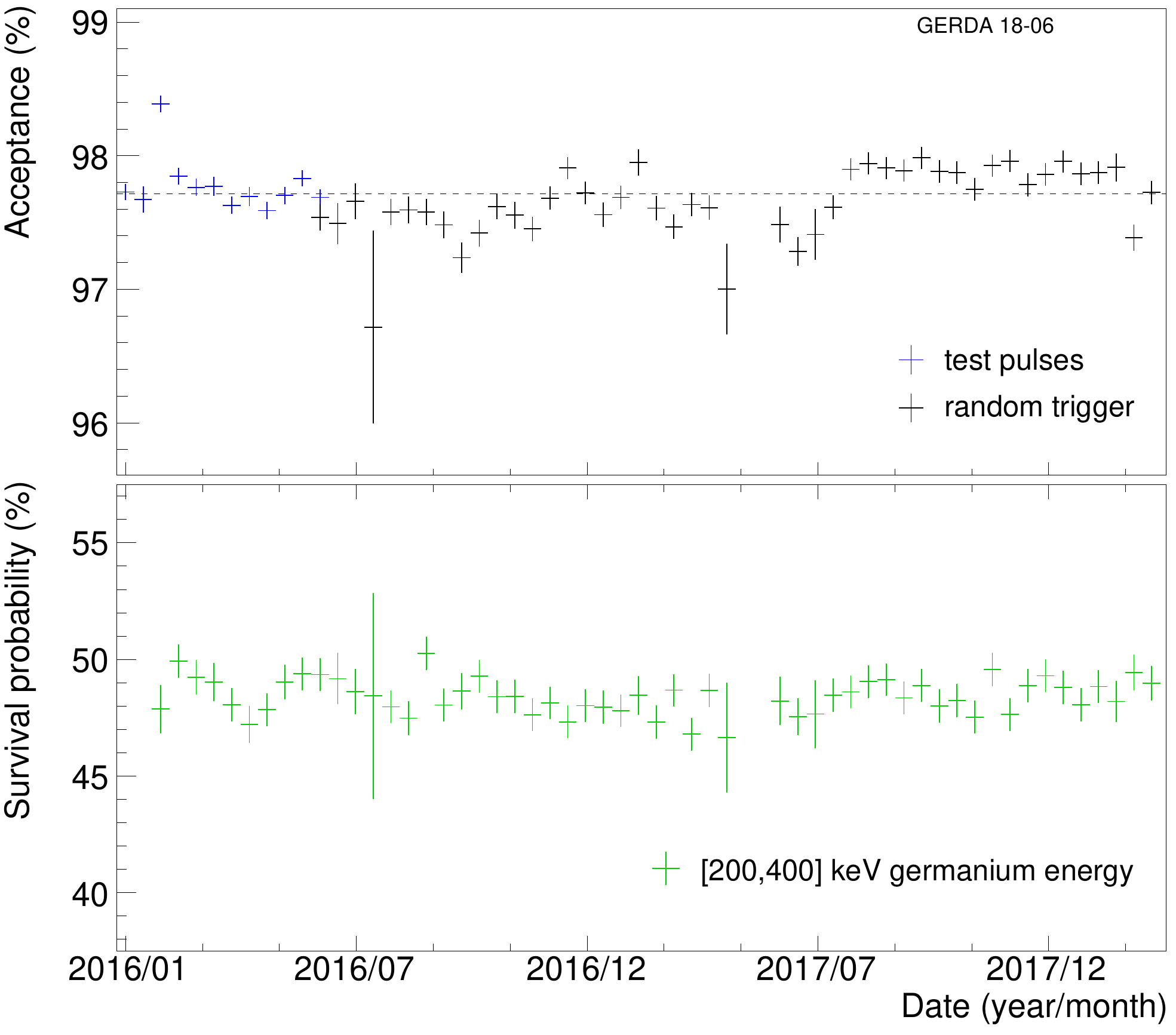}
	\caption{
Stability of signal acceptance and rejection power of the LAr veto. 
The accidental loss is measured by software triggered events.
The rejection is given in terms of survival probability for germanium detector events at low energies dominated by $^{39}$Ar $\beta$ decays.}
	\label{fig:larstab}
\end{figure}

Fig.~\ref{fig:larstab} shows parameters for monitoring the stability of the LAr veto performance during \gerda\ Phase\,II. 
The top panel shows the acceptance for signal events that is measured using the periodically injected test pulses and software triggered events (random trigger).
The exposure weighted deadtime is (2.3 $\pm$ 0.1)\%, predominately caused by random coincidences from $^{39}$Ar decays in the LAr.
The bottom panel shows the survival probability for a sample of events with a signal in the Ge detectors with energy between 200 and 400 keV.
Such events are primarily due to decays of $^{39}$Ar ($Q$-value of 565\,keV) on the Ge detector surface. 
The suppression depends on the detection efficiency of the LAr veto and its stability implies no significant loss in the
rejection power over the full period of data taking.

\begin{figure}[tb]
    \centering
	\includegraphics[width=0.66\textwidth]{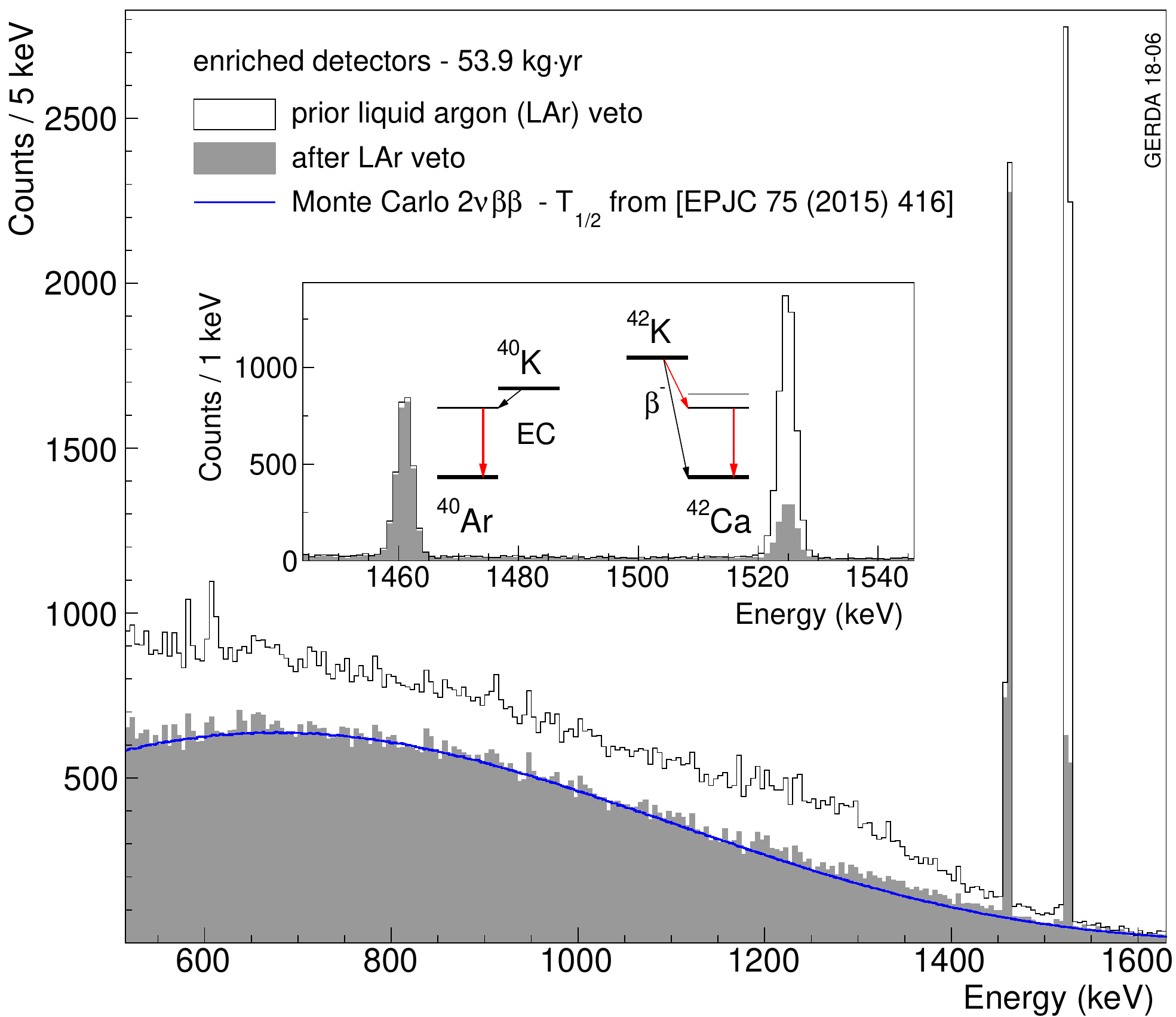}
	\caption{
Spectrum in the \nnbb\ decay dominated energy region before and after LAr veto suppression and
a spectrum of simulated \nnbb\ decays (solid line).
The inset shows a zoom of the full energy $\gamma$ lines from $^{40}$K and $^{42}$K.}
	\label{fig:larspec}
\end{figure}

Fig.~\ref{fig:larspec} shows the impact of the LAr veto cut in the energy region dominated by \nnbb\ decays.
The background due to the Compton scattering of $\gamma$ rays from $^{40}$K and $^{42}$K is cleary visible before LAr veto.
The inset shows a zoom around their full energy peaks (FEPs).
In case of $^{42}$K, it is possible to efficiently veto the $\gamma$ ray event due to an energy deposit in the argon from the 
preceding $\beta$ decay.
For $^{40}$K this is not the case as it undergoes electron capture 
and no energy is available to trigger the LAr instrumentation.
The $^{40}$K FEP is suppressed by (2.5 $\pm$ 0.5)\%, confirming the aforementioned deadtime caused by random coincidences in the LAr veto.

Events from $\gamma$ rays in the Compton continuum below the $^{40}$K and $^{42}$K lines typically feature an energy deposit in the LAr.
Therefore, they are efficiently rejected by the LAr veto and the remainder in the energy spectrum is an almost pure \nnbb\ continuum, 
as indicated by an overlay of the \nnbb\ decay distribution from Monte Carlo simulation normalized using the \nnbb\ decay half-life 
measured during \gerda\ Phase\,I \cite{Agostini:2015nwa}.


\paragraph*{Pulse shape discrimination}

Events from \onbb\ decays are homogeneously distributed 
in the detector volume and the energy deposition is typically contained within
1~mm$^3$ (i.e. single-site events, SSEs). There are two types of background events 
that can be distinguished from these signal-like events using pulse shape 
discrimination (PSD): multiple energy depositions due to $\gamma$ rays (i.e. multiple-site events, MSEs) 
and surface events due to $\alpha$ or $\beta$ decays on the detector n+ or p+ electrode. Energy depositions near 
the n+ electrode of the detector 
can create pulses with long risetimes and incomplete charge collection due to the low electric 
field in the undepleted part of the Li diffusion layer. 
On the other hand, energy depositions near the p+ electrode result in fast 
signals because 
both electrons and holes drift simultaneously through volumes of large weighting potentials; hence the
combined contribution has a shorter rise time.
Since the n+ electrode has a dead surface layer with about 1~mm thickness,
that blocks $\alpha$ particles, they 
can reach the detector active volume only through the much thinner dead layer 
of the p+ electrode and the groove.

The PSD techniques for the rejection of MSEs and surface events applied in Phase\,II 
of \gerda\ are based on the ones used in Phase\,I \cite{Agostini:2013jta} 
with additional improvements. 
    Figs. 2 and 3 of this reference \cite{Agostini:2013jta} show typical examples of pulse shapes from BEGe and 
    coaxial detectors, respectively.
In the case of the BEGe detectors one 
parameter, $A/E$, is used to reject background events, where $A$ is the maximum 
current amplitude and $E$ is the energy, proportional to the area of the current pulse.
As MSEs and n+ surface events are 
characterized by wider current pulses, they feature 
a lower $A/E$ value compared to SSEs, while p+ 
surface events show a higher $A/E$. Therefore rejecting events on both sides of the 
$A/E$ distribution of SSEs enhances the signal to background ratio. Due to their 
different geometry and electric field configuration, the coaxial detectors 
have more complicated patterns of SSE pulses than the BEGe type detectors. Therefore an 
artificial neural network (ANN) is used to discriminate between SSEs and MSEs. 
In addition to these cuts that were already applied in the previous data releases of 
\gerda, a cut on the coaxial detector pulse risetime is now also applied to reject
fast signals from surface events due to $\alpha$ decays near the p+ electrode and in the groove.

The calibration and training of the PSD methods is based on $^{228}$Th calibration 
data. The double escape peak (DEP) at 1593~keV  
is used as a sample of SSEs. The DEP is indeed composed of pair-production events from the 2615~keV $^{208}$Tl line in which the produced electron and positron deposit their 
energy in a single site and the two $\gamma$ rays from the positron annihilation escape the detector 
without further interaction. Similarly, the FEP of the 
$^{212}$Bi decay at 1621~keV is used as a sample of MSEs to train the ANN and to 
test the performance of the $A/E$ method, since $\gamma$ rays are likely to deposit their energy
in multiple Compton scatterings in the detector.

The rejection of p+ surface events is studied with physics data using high energy events
(above the $^{42}$K decay endpoint at $\sim$3500~keV) coming from $\alpha$ decays on the detector surface. 
Additionally, a pure sample of SSEs in the bulk of the detector is provided  by
 \nnbb\ decay events, which compose more than 97\% of the events in the region between 1000-1300~keV after the LAr veto is applied.

The analysis of the raw $A/E$ from calibration data follows similar steps as in 
Phase\,I~\cite{Agostini:2013jta}. The mean $A/E$ in the DEP is scaled to unity in 
stable periods while unstable periods in $A/E$ for specific detectors are excluded 
from the analysis. The slight energy dependence of $A/E$ due to the increase of the 
charge cloud size with energy is also corrected. Because of the higher noise 
in Phase\,II the discrimination threshold on the corrected $A/E$ needs to 
be energy dependent. This is achieved by defining as PSD classifier $\zeta = 
(A/E-1)/\sigma_{A/E}(E)$ where the $A/E$ distribution of SSEs is a Gaussian centered around 1 
with an energy dependent sigma $\sigma_{A/E}(E)$. The low threshold for rejecting 
MSEs is chosen such to have a $90\%$ survival fraction of the DEP. This condition results in cut values on $\zeta$
between $-1.9$ and $-1.2$ depending on the detector. The high 
threshold for rejecting $\alpha$ events is chosen to be at $\zeta=4$ for all BEGe 
detectors. This high cut can be interpreted as a volume cut around the p+ surface 
and the groove of the detectors.

The input variables for the ANN analysis of the coaxial detectors are extracted from the time distribution of 
the signals. After smoothing the pulses, the times
when they reach 1, 3, 5, ..., 99\% of their maximum height are determined using 
linear interpolation between the samples. These 50 times are used as input 
parameters for the multivariate analysis. The network layout consists of two hidden 
layers with 50 and 51 neurons in addition to the 50 input neurons and the one 
classifier output. As described above, DEP events with energy within one FWHM of the 
peak are used as a sample of SSEs and $^{212}$Bi FEP events as MSEs for the 
training. Overtraining was tested and excluded by comparison of the discrimination
sensitivity obtained from the training and independent test samples.
The MSE rejection threshold is chosen at 90\% survival fraction of the 
DEP, resulting in cut values on the classifier 
between 0.302 and 0.455 depending on the detector (c.f. Fig.~\ref{fig:scatter_phy}).

\begin{figure}[ht]
	\centering
    \includegraphics[width=0.75\textwidth]{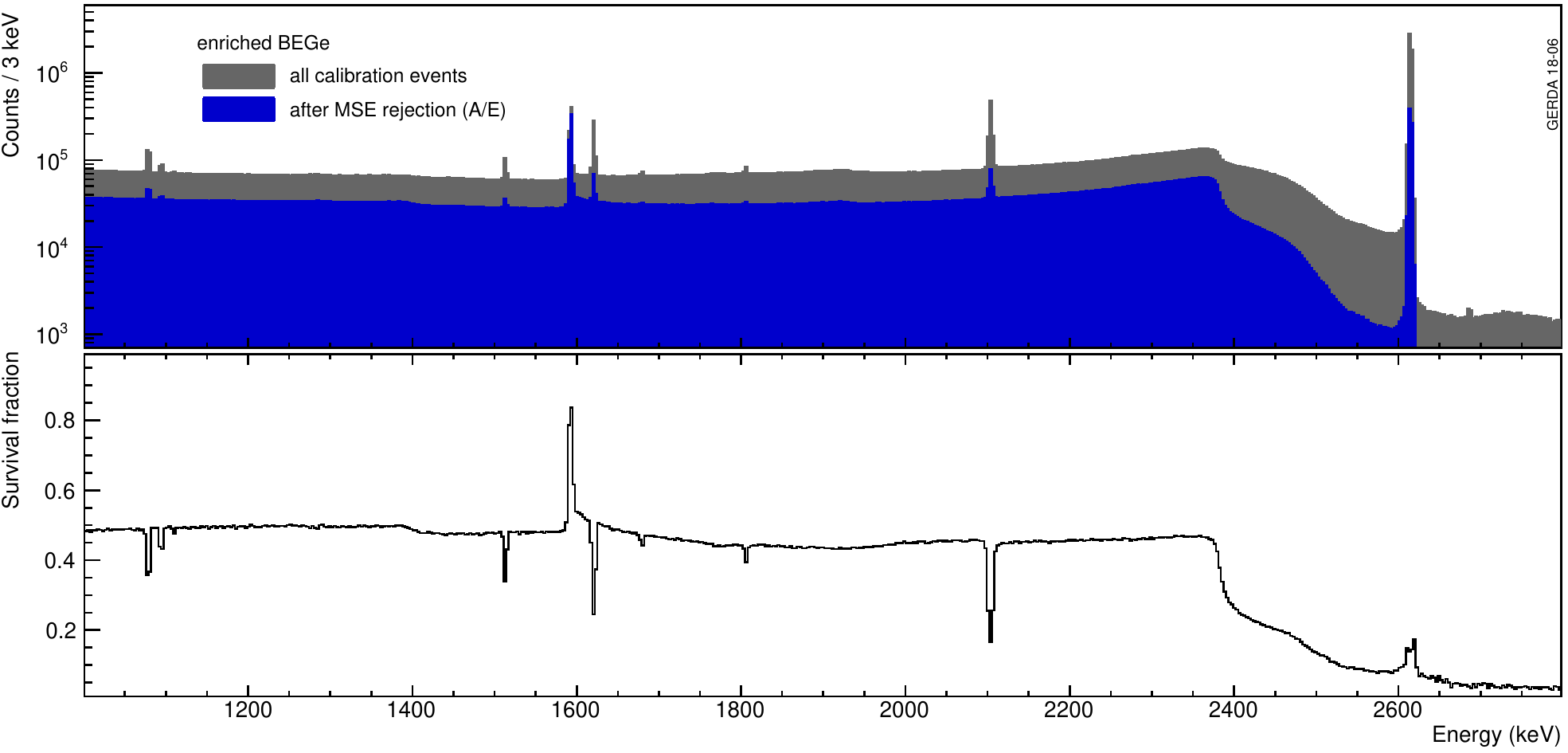}\\
    \includegraphics[width=0.75\textwidth]{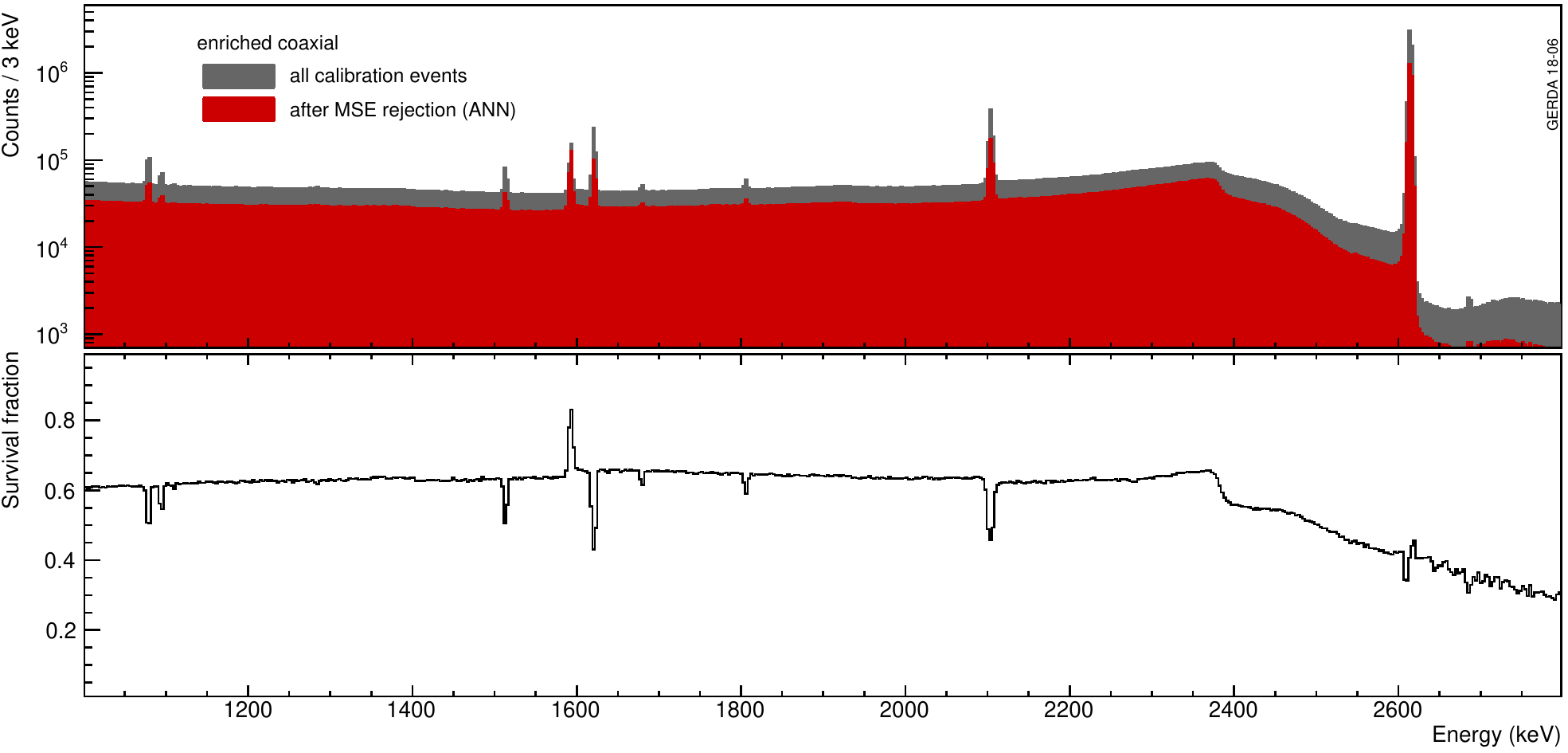}
    \caption{$^{228}$Th calibration spectra of BEGe (top) and coaxial detectors (bottom)  
    before and after multi-site event rejection using the two different PSD 
    techniques. The respective bottom panel shows the survival fraction as a function of 
    energy that is the ratio of the spectra with and without PSD.}
    \label{fig:cal_spectrum_psd}
\end{figure}

\begin{figure}[ht]
	\centering
    \includegraphics[width=0.49\textwidth]{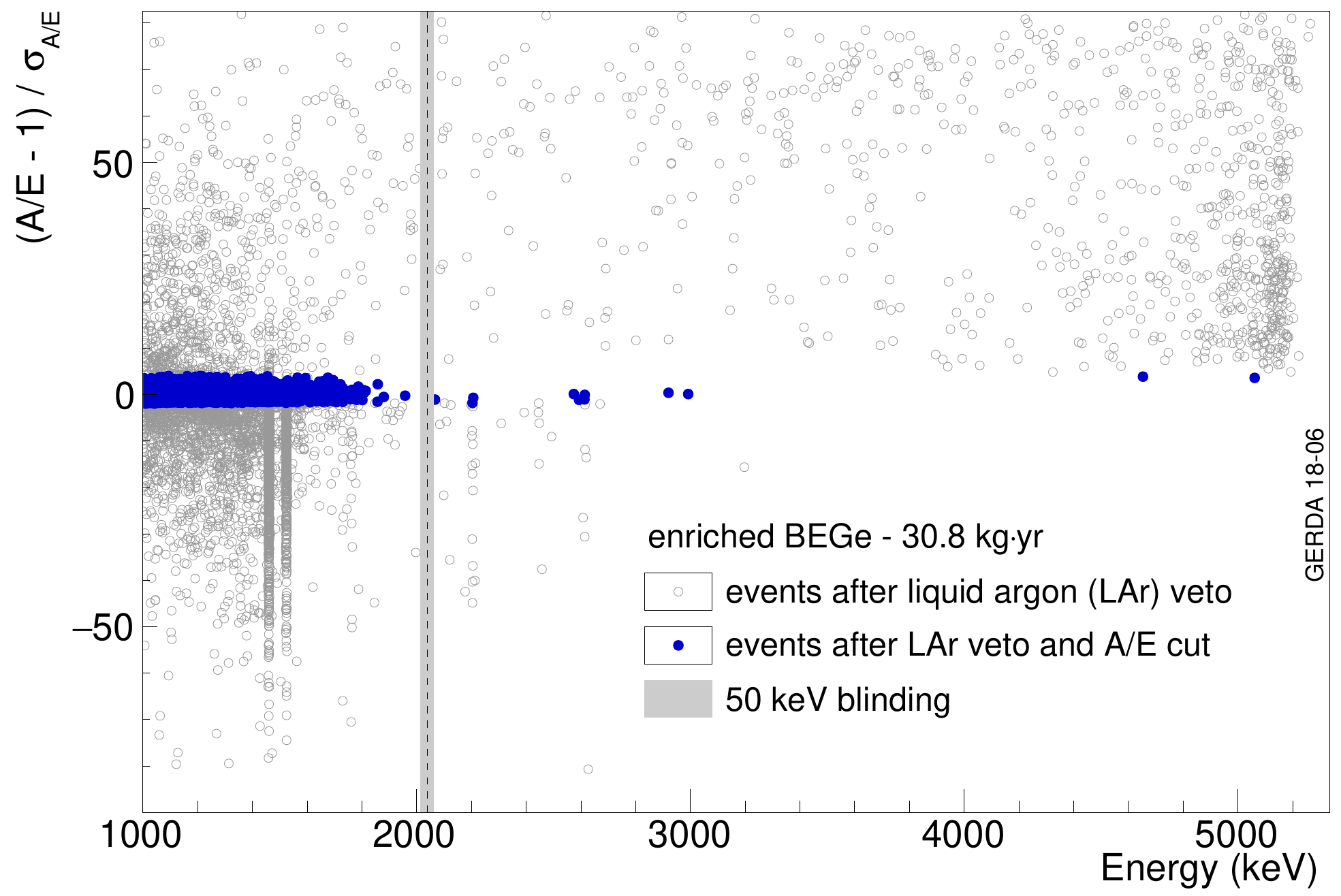}\\
    \includegraphics[width=0.49\textwidth]{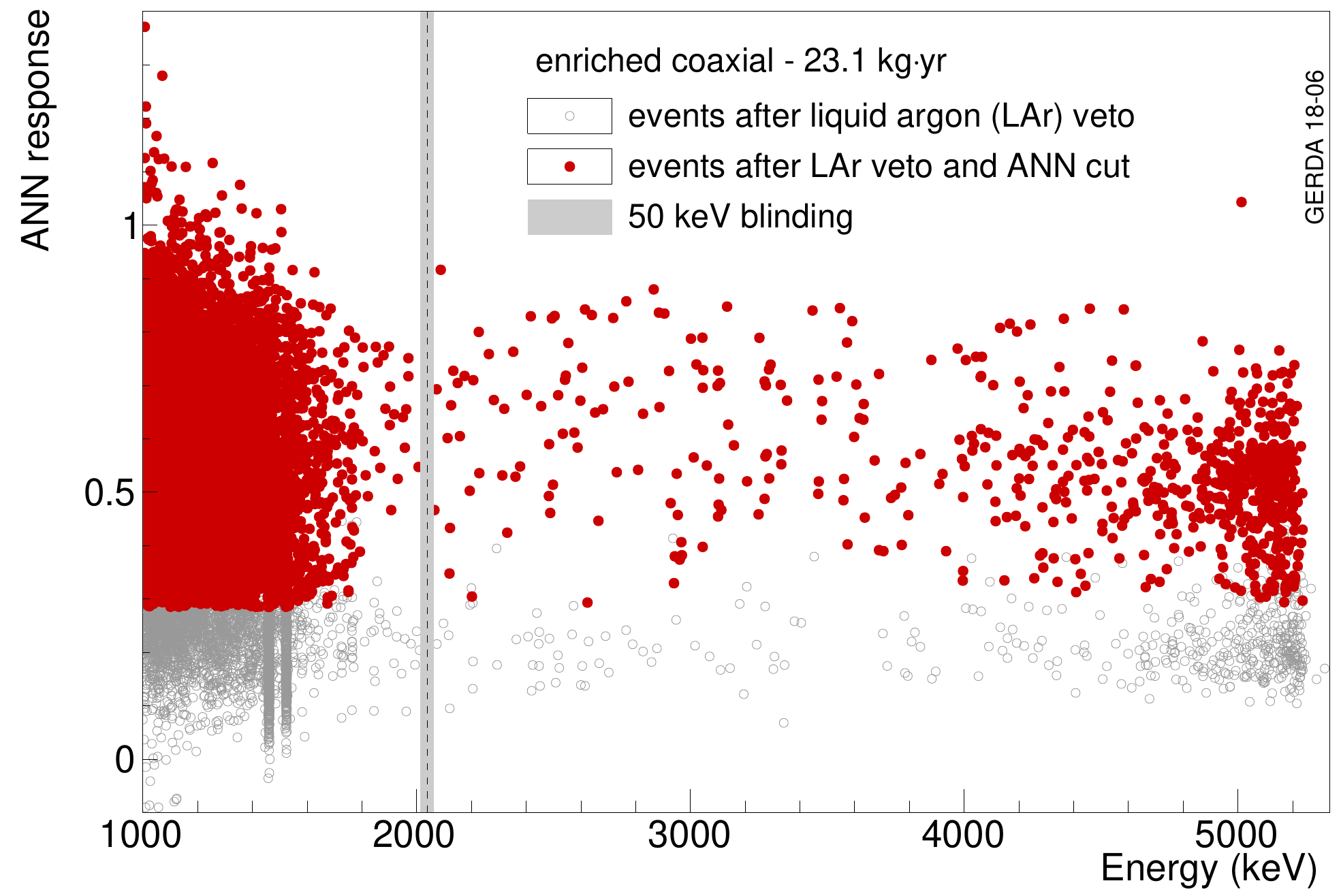}
    \includegraphics[width=0.49\textwidth]{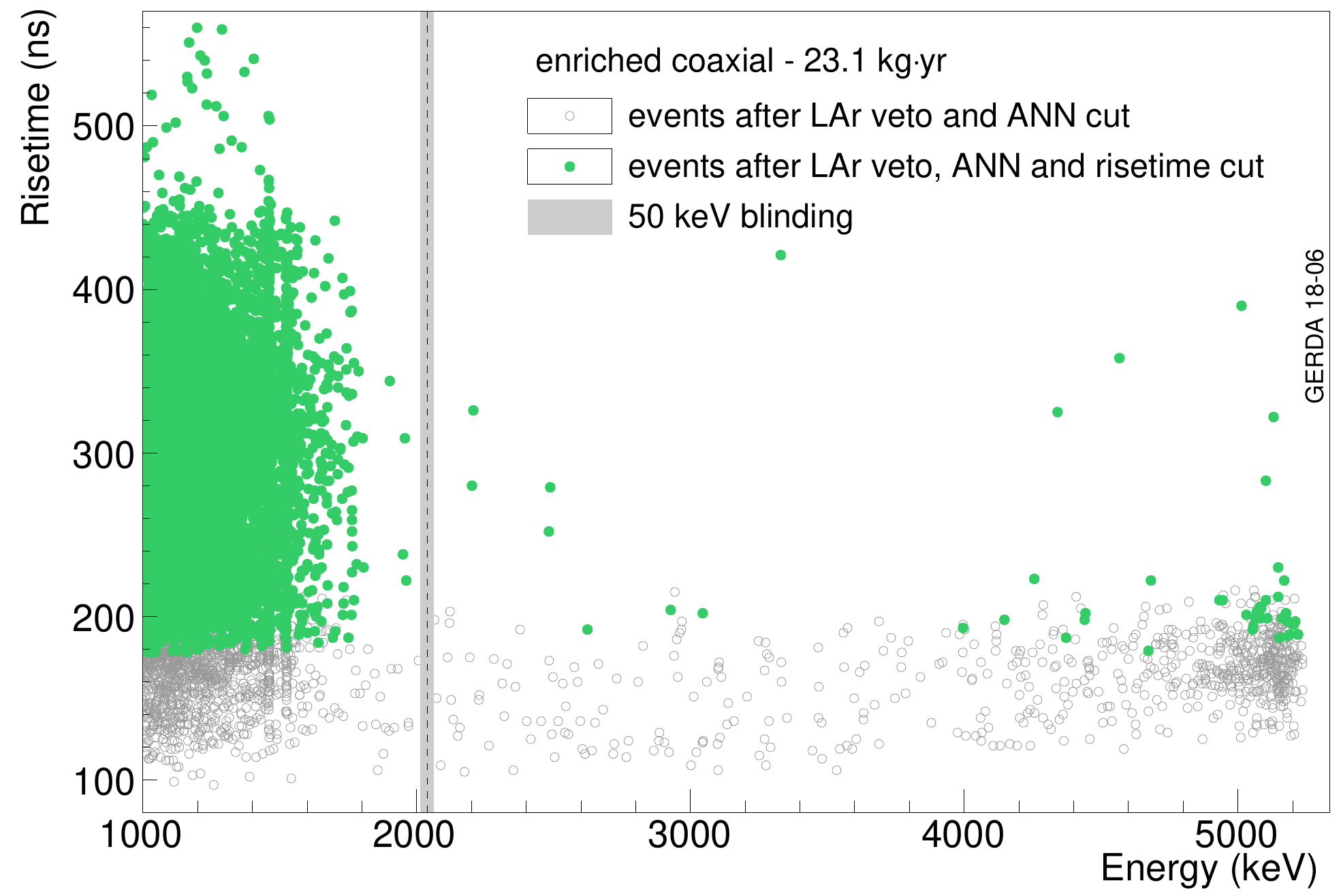}
    \caption{PSD classifiers as a function of energy in physics data after LAr veto 
    from the three different techniques: the blue dots represent events from BEGe 
    detectors surviving the $\zeta$ cut (top), the red ones show events from the 
    coaxial detectors surviving the ANN MSE cut (bottom left), the green ones show 
    events from the coaxial detectors surviving both the ANN MSE and the RT 
    cuts (bottom right).}
    \label{fig:scatter_phy}
\end{figure}

Fig.~\ref{fig:cal_spectrum_psd} shows the $^{228}$Th calibration spectra for BEGe 
and coaxial detectors before and after applying the low side $A/E$ cut and the 
ANN MSE rejection, respectively. The survival fractions in the bottom panels show 
that all $\gamma$ lines are more strongly suppressed than the DEP, and the suppression of the Compton 
continuum is relatively independent of the energy up to the Compton edge.

The risetime (RT) of the pulses used for rejecting surface events in
coaxial detectors is extracted after noise reduction and interpolation of the digital signals. 
The RT parameter is defined as the
difference between the time at which the signal is at 10\% and 90\% of its maximum amplitude.
This parameter has been found to be robust against noise and other changes in detector 
conditions. The RT threshold is determined by maximizing the acceptance of 
\nnbb\ decay events ($\varepsilon_{\nnbb}$) and at the same time 
minimizing the survival fraction of surface $\alpha$ events with energy above 
3500~keV ($\varepsilon_{\alpha}$). The figure of merit chosen for the optimization 
is $f(t)=\varepsilon_{\nnbb}^2(t)\cdot(1-\varepsilon_{\alpha}(t))$ where 
$t$ is the RT threshold. This optimization is performed on physics events that 
survive the MSE rejection because preliminary pulse shape simulations have shown 
that the ANN and the RT cut are rejecting events from complementary regions near 
the detector surface. The ANN classifies events near the borehole surface as MSEs 
while the RT cut removes fast events originating from the regions around the 
bottom of the borehole and the groove.

Fig.~\ref{fig:scatter_phy} shows the physics data after LAr veto with and without 
the PSD selections for the three different techniques. The $A/E$ and the ANN both 
suppress the FEPs from $^{40}$K and $^{42}$K decays as expected. Furthermore, all 
high energy $\alpha$ events have relatively high $A/E$ values and almost all are rejected.
The combination of the ANN and RT 
cuts in the case of the coaxial detectors rejects the high energy events from 
surface $\alpha$ decays with about 96\% probability.

The \onbb\ decay survival fraction and its uncertainty is estimated for the different PSD 
techniques in various ways. In the case of BEGe detectors, the DEP survival fraction 
is assumed as an estimate of the \onbb\ survival fraction. Differences between the expected 
\onbb\ decay signal and DEP events are taken into account in the 
systematic uncertainties: DEP events are more likely to occur close to the sides 
and edges of the detector while \onbb\ decay events are distributed 
equally in the detector volume, this introduces about 1\% difference based on 
\textsc{Geant4} simulations. Additionally, because of its higher energy, the 
\onbb\ decay signal has a higher fraction of MSEs that introduces an 
additional 2\% uncertainty based on \textsc{Geant4} simulations. Finally, the uncertainty of 
the DEP survival fraction is estimated by shifting the low $A/E$ threshold by its 
uncertainty. This is based on the agreement of $A/E$ distributions between physics 
and calibration data and on differences between methods to determine the cut 
value. The threshold has been shifted by 0.2 resulting in a 1.7\% relative 
uncertainty on the DEP survival fraction. The estimated \onbb\ survival fraction of the BEGe data set 
is hence $(87.6\pm0.1\mathrm{(stat)}\pm2.5\mathrm{(syst)})\%$.
\begin{figure}[ht]
        \centering
	    \includegraphics[width=0.55\textwidth]{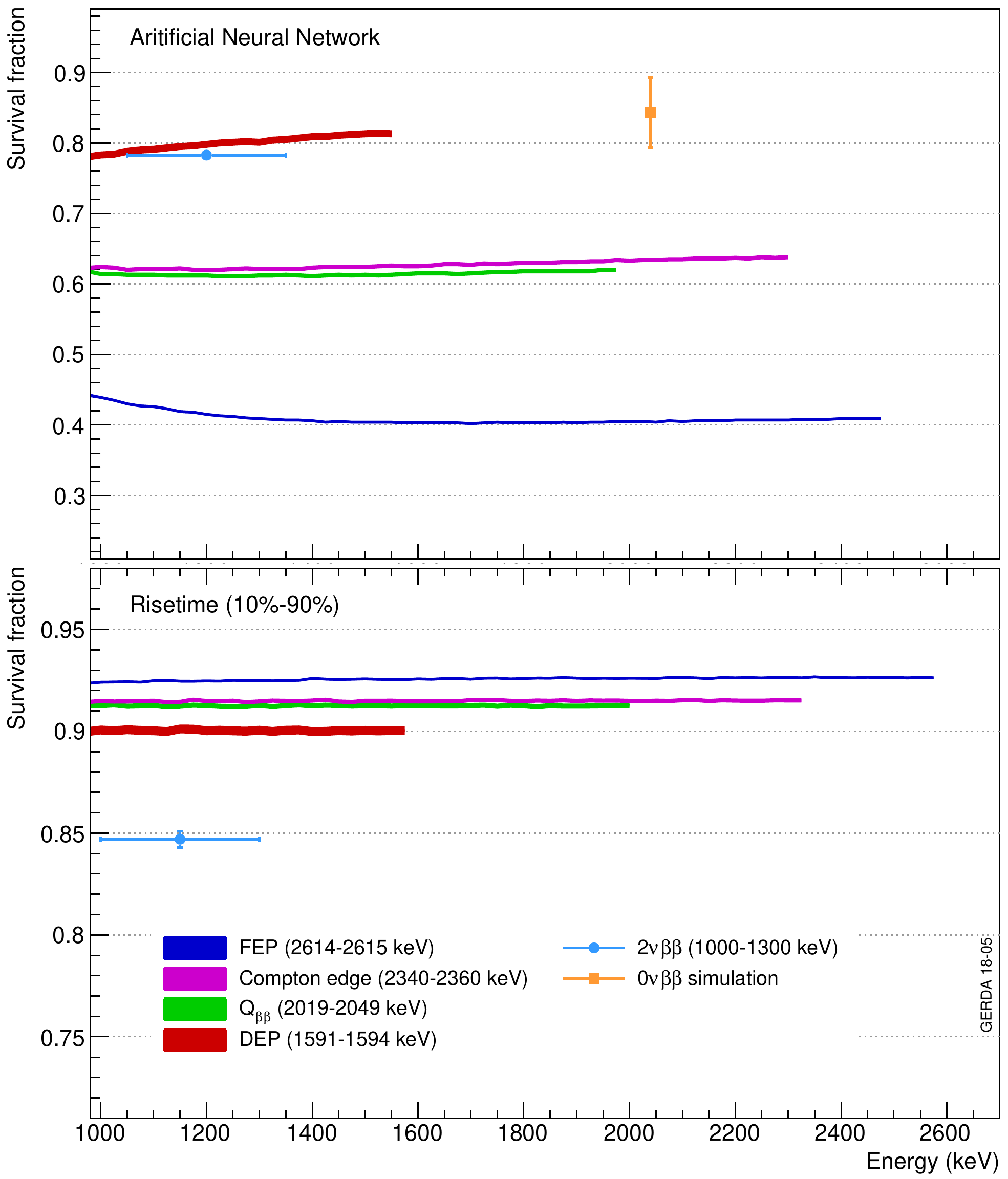}
	        \caption{For coaxial detectors the survival fraction of various event samples from calibration data with
		    rescaled noise corresponding to signals with different energies together with
		        the survival fraction of \nnbb\ decay events from physics data. The
			    top panel shows the results from the ANN MSE rejection and the bottom panel from
			        the surface event rejection using the RT cut.}
				    \label{fig:psd_energydep}
				    \end{figure}

In the case of the coaxial detectors the \onbb\ decay survival fraction
has been estimated using pulse shape simulations and \nnbb\ 
decay events, taking into account the energy dependence of the PSD techniques. 
The survival fraction of peaks in the calibration data and of the 
\nnbb\ decay events in physics data has been reproduced with 
pulse shape simulations within a few percent. This gives confidence in the 
survival fraction of \onbb\ decay signal estimated from the simulations that 
is found to be (84 $\pm$ 5)\%. The uncertainty has been estimated studying the energy dependence with event samples 
from calibration data with rescaled noise corresponding to lower effective 
energies. The survival fraction of the different event samples as a function of 
this effective energy is shown in Fig.~\ref{fig:psd_energydep} for both the ANN 
and the RT cuts. The ANN shows a slightly different energy dependence for each event sample, suggesting 
that the cut does not only depend on the noise but also on the topology of the event 
sample. Using the slopes from the different event samples to extrapolate the 
survival fraction from the \nnbb\ decay survival fraction gives a 
conservative systematic uncertainty of $\pm5\%$. In the case of the RT cut the
survival fraction shows no significant energy dependence for all type of
events considered. This allows to estimate
the \onbb\ decay survival fraction directly from the sample of \nnbb\ decay events that gives
$(84.7\pm0.4\mathrm{(stat)}\pm1.0\mathrm{(syst)})\%$, where the systematic uncertainty accounts for minor differences between the two samples.
The final survival fraction for the new data set
of the coaxial detectors is hence given by the combination of the two PSD cuts:
$(71.2\pm4.3)\%$.

\begin{figure}[ht]
	\centering
    \includegraphics[width=0.75\textwidth]{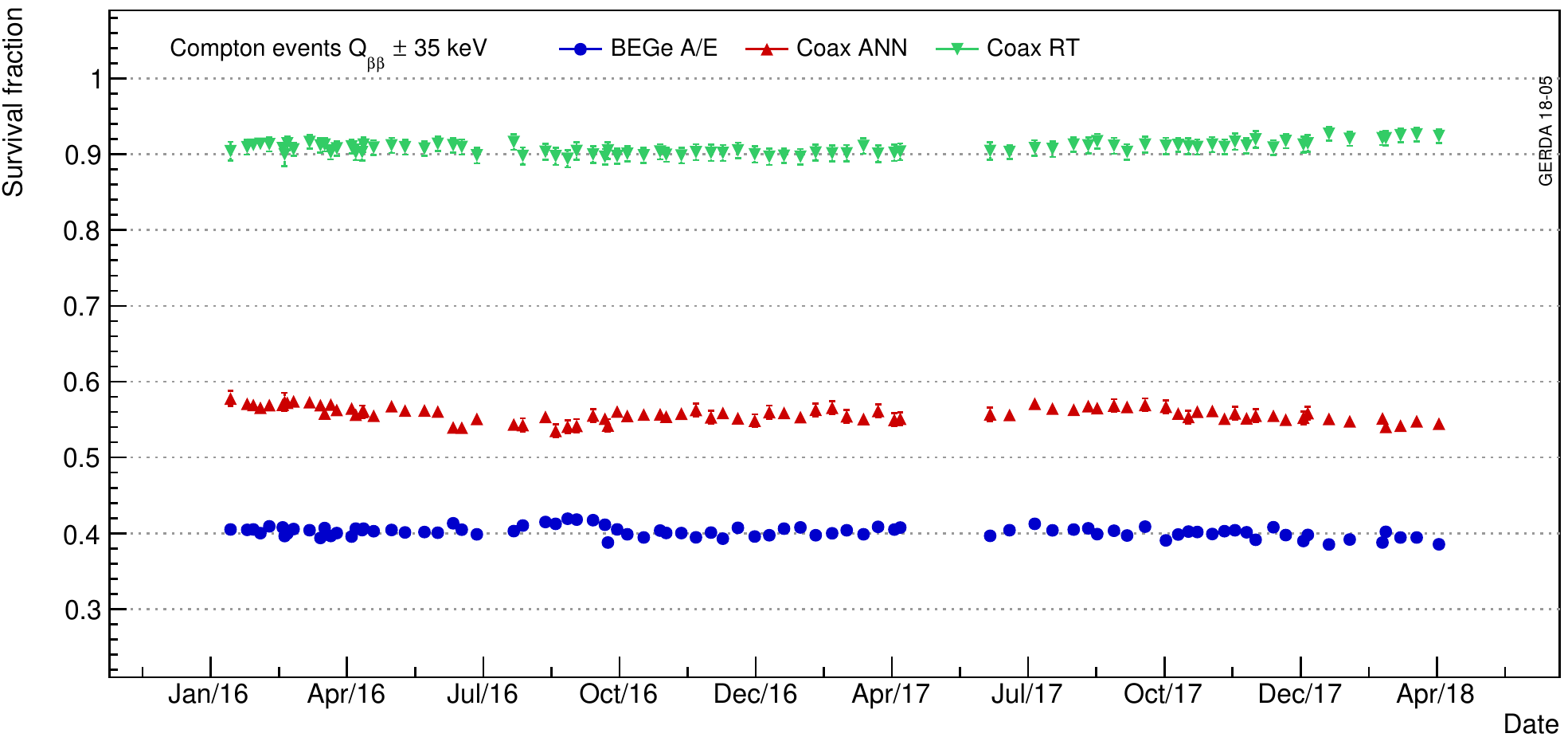}
    \caption{Survival fraction of Compton events with energy within 
    $\qbb\pm35\mathrm{~keV}$ from the $^{228}$Th calibrations throughout the data 
    taking demonstrating the stability of the PSD techniques.}
    \label{fig:psd_stability}
\end{figure}

The stability in time of the survival fraction of the different PSD techniques has also been 
studied and is shown in Fig.~\ref{fig:psd_stability}. The suppression factor of 
background Compton events in the region of interest is stable at a 2\% level over 
more than two years of data taking. 

An additional pulse shape analysis cut has been introduced for both, BEGe and coaxial detectors, in order to reject events
with slow or incomplete charge collection from external background on the n+ surface of the 
detectors.
The energy of these events is biased 
because the ZAC energy reconstruction algorithm uses a finite integration time. 
The cut is based on 
the difference between two reconstructed energy values using the same method but 
different integration times.
The threshold is set to about 0.25\% difference in reconstructed energy 
between the two estimators with different integration times,
that corresponds to about 5~keV at \qbb.
The survival fraction of the cut is 99.8\% for signal events.


\paragraph*{Statistical analysis}
Neutrinoless double-$\beta$ decay is searched for by performing an extended unbinned 
maximum likelihood fit of the energy spectrum. The function used for the fit is the sum of a 
Gaussian distribution centered at \qbb\ (signal term) and a flat distribution, which accounts for 
the constant background (background term). The width of the Gaussian term is determined by 
the energy resolution, while its amplitude depends on the half-life \thalfzero\ of \onbb\ 
decay. 

Following the same strategy of refs.~\cite{gerdaPRL2013,nature544,Agostini:2018tnm}, detectors 
having similar performance and similar background level are grouped together and their data treated as 
a single data set. The number of expected \onbb\ events in each data set is calculated as
\begin{equation}
\mu_S = \ln 2 \cdot \frac{N_A}{m_a}  \cdot \epsilon \cdot \mathcal{E} \cdot \frac{1}{T^{0\nu}_{1/2}}
\end{equation} 
where $N_A$ is the Avogadro number, $m_a$ the molar mass, $\mathcal{E}$ the exposure and $\epsilon$ the 
total selection efficiency. The efficiency $\epsilon$ accounts for: the fraction of 
\gess\ in the detectors (\fgesix\ of Table~\ref{table:detproperties}); the fraction of active volume; 
the probability that \onbb\ events in the active volume release their entire energy \qbb\ inside the 
detectors ($\epsilon_{fep}$ of Table~\ref{table:detproperties}); the efficiency of all reconstruction 
and analysis cuts. The energy resolution and efficiency of each data set are calculated as
exposure-weighted averages of the corresponding values for the individual 
detectors. 

The seven data sets used for the analysis are summarized in Table~\ref{tab:datasets}. 
Data from Phase\,I of \gerda\ 
and from the first part of Phase\,II have been already reported in
refs.~\cite{gerdaPRL2013,nature544,Agostini:2018tnm} and their analysis -- in particular the
PSD classification -- is unchanged. Coaxial detector data from
Phase\,II are split into two data sets because the new PSD
method introduced for the first time in this work provides a significantly lower
background level. All BEGe detector data from Phase\,II are instead in the
same data set as the analysis strategy is consistent with what was previously applied.
In total, in this work we release 23.1\,\kgyr\ of coaxial detector data
(PhaseII-Coax2) and 18.2\,\kgyr\ of BEGe detector data (part of PhaseII-BEGe).

The complete form of the  likelihood function is reported in ref.~\cite{nature544}.
The background rates are free parameters, independent for each data set, while the 
signal strength $S = 1/T^{0\nu}_{1/2}$ is a free parameter which is in common to 
all data sets. All parameters are bound to positive values.
The fit is performed in the energy range between 1930 and 2190 keV, except 
the windows $(2104 \pm 5)$~keV and $(2119 \pm 5)$~keV where known peaks are
expected. The total width of the analysis window is 240~keV. 
\begin{table}[tbp]
\centering
\caption{Summary of the Phase\,I and Phase\,II analysis data sets: exposure \exposure, energy 
resolution (FWHM) at \qbb, total efficiency $\epsilon$, background rate B, and 
number $N$ of events in the 240\,keV analysis window. 
Note that the BI is computed excluding the events in a 10\,keV window centered at the \qbb.\label{tab:datasets}}
\vspace*{5mm}
\begin{tabular}{lc c c c c}
\hline
\hline \\[-2ex]
Data set & $\mathcal{E}$ & FWHM & $\epsilon$ & B & N \\
 & (\kgyr) & (keV) & & cts/(keV$\cdot$t$\cdot$yr) & \\
\hline
PhaseI-Golden & 17.9 & 4.3(1) & 0.57(3) & $11 \pm 2$ & 46 \\
PhaseI-Silver & 1.3 & 4.3(1) & 0.57(3) & $30 \pm 10$ & 10 \\ 
PhaseI-BEGe   & 2.4 & 2.7(2) & 0.66(2) & $5^{+4}_{-3}$ & 3 \\
PhaseI-Extra & 1.9 & 4.2(2) & 0.58(4)  & $5^{+4}_{-3}$ & 2 \\
\hline
PhaseII-Coax1 & 5.0 & 3.6(1) & 0.52(4) & $3.5^{+2.1}_{-1.5}$ & 4 \\
PhaseII-Coax2 & 23.1 & 3.6(1) & 0.48(4) & $0.6^{+0.4}_{-0.3}$ & 3 \\
PhaseII-BEGe & 30.8 & 3.0(1) & 0.60(2) & $0.6^{+0.4}_{-0.3}$ & 5 \\
\hline
\hline
\end{tabular}
\end{table}
\begin{table}[tbp]
\centering
\caption{List of Phase\,II events in the analysis window 1930-2190~keV, which survive 
all selection cuts (anti-coincidence, muon veto, LAr veto, PSD) \label{table:eventtable}.}
\vspace*{5mm}
\begin{tabular}{l c c c}
\hline
\hline \\[-2ex]
Data set & Energy & Date/time & Detector \\
 & (keV) & (UTC) & \\
\hline 
PhaseII-Coax1 & 1995.2 & 10 Feb 2016 13:04 & ANG~4 \\
PhaseII-Coax1 & 1968.0 & 13 Mar 2016 04:42 & ANG~3 \\
PhaseII-BEGe & 1958.7 & 13 Mar 2016 05:40 & GD61C \\
PhaseII-Coax1 & 2063.6 & 28 Mar 2016 16:00 & ANG~3 \\
PhaseII-Coax1 & 2060.6 & 22 May 2016 11:44 & ANG~1 \\
PhaseII-BEGe & 2018.1 & 30 Aug 2016 01:57 & GD35B \\
PhaseII-Coax2 & 1950.9 &  9 Oct 2016 02:44 & ANG~1 \\
PhaseII-BEGe & 2068.0 & 27 Nov 2016 23:47 & GD35B \\
PhaseII-BEGe & 2056.4 & 31 Jan 2017 07:48 & GD91A \\
PhaseII-BEGe & 2042.1 & 24 Aug 2017 12:48 & GD76C \\
PhaseII-Coax2 & 1962.7 &  1 Nov 2017 01:42 & ANG~1 \\
PhaseII-Coax2 & 1957.5 & 16 Jan 2018 22:46 & RG~1 \\ 
\hline 
\hline
\end{tabular}
\end{table}

The events of Phase\,II between 1930 and 2190~keV which survive all 
selection cuts are listed in Table~\ref{table:eventtable}.
The likelihood of the \gerda\ data is maximized for null signal strength, i.e. the 
best-fit is $S=1/T_{1/2}=0$. The confidence (most probable) interval on $S$ is 
then evaluated in the frequentist (Bayesian) framework, as described 
in detail in ref.~\cite{nature544}.

The frequentist analysis makes use of a two-sided test statistics based 
on the profile likelihood $\lambda(S)$, in which all background rates 
are treated as nuisance parameters. 
Systematic uncertainties are accounted in the fit by additional nuisance parameters 
constrained by Gaussian pull terms. They are the uncertainties on efficiency and 
energy resolution for each data set, as reported in Table~\ref{tab:datasets}, and a possible 
common systematic shift of the energy scale, which is modeled as Gaussian having zero mean 
and 0.2~keV standard deviation.
As Wilk's approximation does not hold in the low-statistics regime of \gerda, 
a Neyman construction of the confidence interval is performed through a toy
Monte Carlo on the conditional parameter space of the best fit on data.
The derived confidence interval of $S$ is:
\begin{equation}
T_{1/2} > 0.9 \cdot 10^{26} \ \textrm{yr (90\% C.L.).}
\end{equation}
The systematic uncertainties have a marginal impact and weaken the limit by 
less than 1\%.
The median sensitivity for limit setting, i.e. the median expectation for 
the frequentist limit, is evaluated by an ensemble of Monte Carlo 
generated data sets, having the same parameters of \gerda\ and null 
injected signal ($S=0$). The calculation carried out with the parameters of 
Table~\ref{tab:datasets} gives a sensitivity of $1.1 \cdot 10^{26}$~yr, 
thus making \gerda\ the first experiment to surpass $10^{26}$~yr. The actual 
realization of \gerda\ is slightly below the median sensitivity: the p-value, 
i.e. the probability to obtain a limit stronger than the actual one in an 
ensemble of repeated experiments with null signal, is 63\%. \\

\begin{figure}[htb]
\centering
\includegraphics[width=0.8\textwidth]{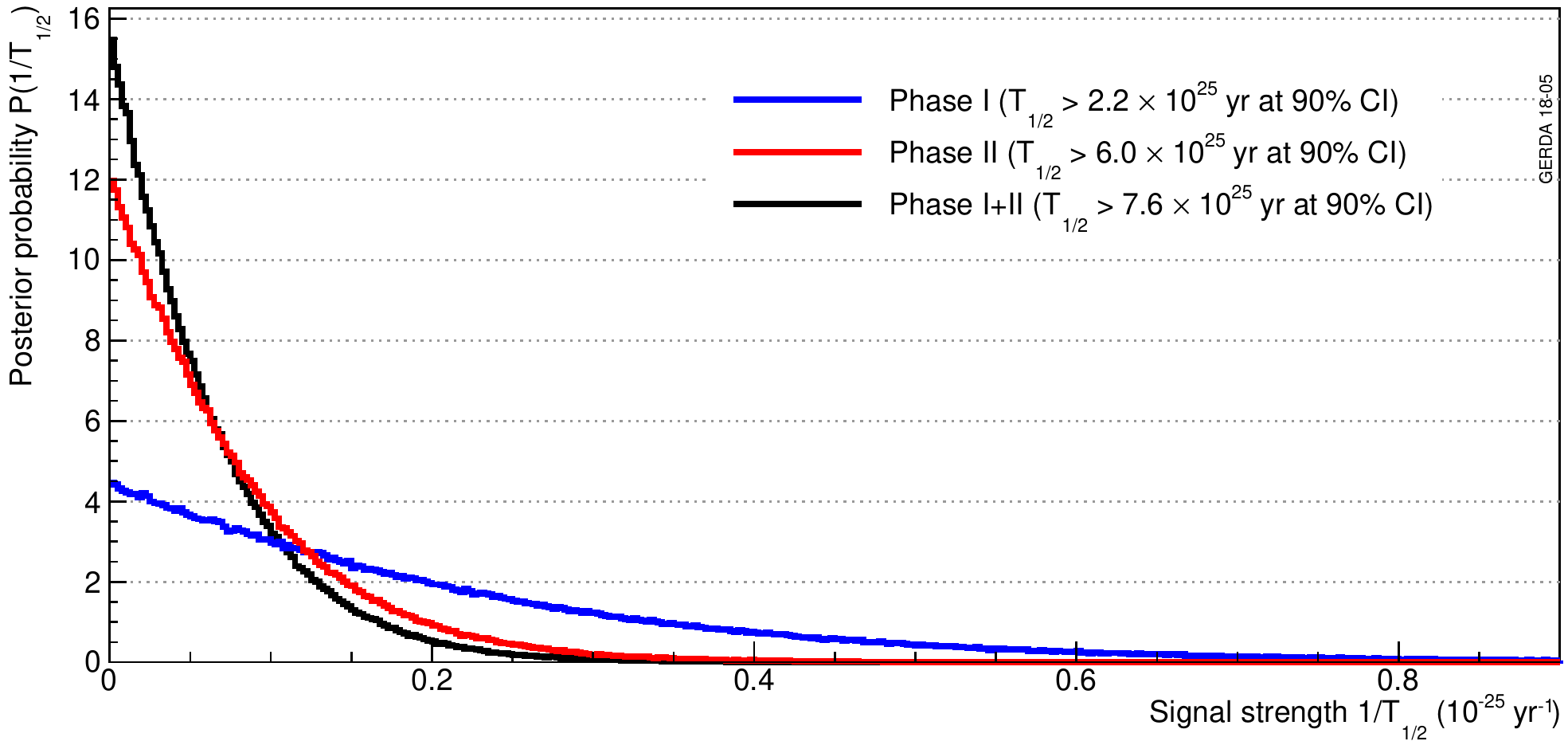}
\caption{Marginalized posterior probability density $P(S)$ for the \onbb\ signal strength from the 
analysis of Phase~I only (four data sets, blue), Phase~II only (three data sets, red) and 
the combination (seven data sets, black). In all cases, the prior distribution $P_0(S)$ is 
constant between 0 and $10^{-24}$ yr$^{-1}$.}
\label{fig:bayes} 
\end{figure}

The statistical analysis is also performed within a Bayesian framework. The
full posterior probability density function (PDF) is calculated from the seven 
data sets according to Bayes’ theorem. The posterior PDF on the signal 
strength is then obtained by marginalization over all nuisance 
parameters, via a Markov chain Monte Carlo numerical integration, as described 
in ref.~\cite{nature544}. The prior PDF for all background indices is taken to 
be constant between 0 and 0.1~\ctsper; as in previous 
work~\cite{gerdaPRL2013,nature544,Agostini:2018tnm}, a flat prior distribution 
between 0 and $10^{-24}$ yr$^{-1}$ is taken as the prior $P_0(S)$ for $S$,
namely all counting rates up to a 
maximum are considered to be equiprobable. Systematic uncertainties are 
folded into the posterior PDF as integral averages~\cite{nature544}, with the integration 
being performed numerically by a Metropolis algorithm. As in the 
frequentist case, the impact of systematic uncertainties is at the 
percent level. The marginalized posterior PDF $P(S)$ is displayed in 
Fig.~\ref{fig:bayes} for Phase~I only, Phase~II only and the combination of the two. 
The 90\% credible interval (CI) for the half-life derived from the full data set is 
\begin{equation}
T_{1/2} > 0.8 \cdot 10^{26} \ \textrm{yr (90\% CI).}
\end{equation} 
The median sensitivity of the experiment in the case of no signal is 
$0.8 \cdot 10^{26}$~yr. Also in the Bayesian framework, the real
realization of \gerda\ is slightly below the median sensitivity: the 
probability to obtain a stronger limit is 59\%. 

The Bayesian analysis can be also performed with Phase~II data only, using 
the posterior PDF from Phase~I as the prior $P_0(S)$ on $S$: this procedure yields 
exactly the same result as the combined analysis of the seven data sets of 
Table~\ref{tab:datasets}. As discussed in ref.~\cite{nature544}, other  
reasonable choices for the prior distribution $P_0(S)$ typically result in more 
stringent limits with respect to the flat prior in $S$ considered above. For 
instance, the assumption of equiprobable effective neutrino masses, i.e. 
$P_0(S) \propto 1/\sqrt{S}$, gives a 90\% credible interval 
$T_{1/2} > 1.2 \cdot 10^{26}$~yr.

The fact that the actual \gerda\ data yield a limit weaker than the median
sensitivity in both frequentist and Bayesian analysis is due to the presence of
an event close to \qbb\ in the PhaseII-BEGe data set where the background
expectation is about 0.01\,cts/keV.
This event has an energy of 2042.1~keV and is 2.4\,$\sigma$ away from \qbb. 
Taking into account the statistical uncertainties on the estimates 
of the background indices, the probability to observe one or more 
background events in the range $\qbb \pm 2 \sigma$ is about 29\% considering all
data sets of \gerda\ Phase\,II.  

Although the event in the signal region weakens the limit, the statistical
analysis attributes it to the background.
The global \gerda\ likelihood is indeed maximized for no signal counts. This
means that the occurrence of such a background overfluctuation 
is more likely than the occurrence of a single signal event at 2.4\,$\sigma$.
%
The interpretation of this event as due to the background is further 
corroborated by the fact that the detector in which the event 
occurred (GD76C) has a slightly better energy resolution than the average of 
the data set, which is used in the statistical analysis.


\paragraph*{Combination of sensitivities}
We analyze our data using frequentist and Bayesian statistical methods \cite{PDG}.
In the Bayesian case, a prior probability distribution for the half-life is updated
using information from the data to yield a posterior distribution. Hence a direct
measurement/limit of $T_{1/2}$ is the result - however depending on the somewhat
arbitrary choice of the prior. In the frequentist analysis no direct statement on
$T_{1/2}$ can be inferred. Instead, for all true $T_{1/2}$ values in the extracted
(90\%) confidence interval, the chance that an experiment sees a signal strength
(or best-fit $T_{1/2}$) like in our data is 90\%. 
The meaning of this frequentist statement is often ignored and the confidence interval 
is interpreted as an exclusion of model parameters. Because of the frequentist 
construction, data fluctuations can lead to surprising features:
improving the experiment by increase
of exposure might lead to poorer limits on $T_{1/2}$ like it happened for EXO-200
\cite{exo}, or the limit is much stronger than the sensitivity of the experiment
- the median limit on $T_{1/2}$ expected from many repetitions of the experiment
assuming no \onbb \ decay signal - like it is the case for CUORE \cite{cuore} and
all KamLAND-Zen results
(see \cite{kamlzen} and refs. therein).
For the comparison and combination of experiments we therefore propose
to use instead of frequentist limits the sensitivities until there is some hint for a signal.
Bayesian limits are not available for all experiments listed in Table~1.

Assuming exclusively light Majorana neutrino exchange, the relation between
the half-life for neutrinoless double-$\beta$ decay $T_{1/2}$ and the effective 
Majorana mass $m_{\beta\beta}$ is
\begin{equation}
 (T_{1/2})^{-1} =  {\cal G} \cdot g_A^4 \cdot |{\cal M}|^2 \cdot {m_{\beta\beta}}^2 / {m_e}^2
\label{eq:eqsens}
\end{equation}
where $\cal G$ is a phase space factor, $g_A$ the axial vector coupling constant, 
$\cal M$ the nuclear matrix element, and $m_e$ the mass of the electron. 
The phase space factors have been calculated recently with high precision (see 
Table~\ref{tab:phasp_G}).  The nuclear matrix elements (NME), however, exhibit 
a large scatter depending on the chosen model calculation (see Table~\ref{tab:nme_M}). 
\begin{table}[htb]
\begin{center}
\caption{Phase space factors $\cal G$ from two recent calculations \cite{kotila,stefanik}.
\label{tab:phasp_G}
}
\vspace*{5mm}
\begin{tabular}{rc}
\hline
\hline \\ [-2ex]
Isotope &   ${\cal G}~(\times 10^{-15}yr^{-1}$)  \\
\hline \\[-2ex]
$^{76}$Ge &  2.363~~~~~2.37  \\
$^{82}$Se &  10.16~~~~10.18 \\
$^{130}$Te & 14.22~~~~14.25 \\
$^{136}$Xe & 14.58~~~~14.62 \\
\hline
\hline
\end{tabular}
\end{center}
\end{table} 
Hence the combination of the sensitivities from the experiments listed in 
Table~1 has been 
evaluated for each set of NME calculations separately assuming implicitly that each set provides for each isotope 
the correct NME. 
The $T_{1/2}$ sensitivity of each experiment is converted 
into an equivalent  $T_{1/2}$ sensitivity for $^{76}$Ge according to eq.~\ref{eq:eqsens}. Recognizing that most experiments operate in
the background dominated regime where sensitivity scales with the square root of exposure, i.e. the
number of Gaussian fluctuating background counts, we calculate the combined sensitivity $\cal S$$^{comb}$ 
as the square root of the sum of squares of these converted sensitivities.
\begin{table}[htb]
\begin{center}
\caption{
Nuclear matrix elements $\cal M$ calculated by different model calculations
for indicated isotopes using the axial vector coupling constant $g_A$. Combined
effective Majorana masses $m_{\beta\beta}^{comb}$ 
are calculated for each set of nuclear matrix element calculations from the sensitivities of the 
experiments listed in Table~1. 
\label{tab:nme_M}
}
\vspace*{5mm}
\begin{tabular}{lcccccr}
\hline
\hline \\ [-2ex]
Model & $g_A$ &           &    ~~~~$\cal M$   &            &            & $m_{\beta\beta}^{comb}$\\
\cline{3-6} \\ [-2ex]
       &       & $^{76}$Ge & $^{82}$Se & $^{130}$Te & $^{136}$Xe & meV~\\
\hline
SM St-M \cite{smst}  & 1.25  & 2.81  & 2.64 & 2.65   & 2.19   & 143~ \\
SM Mi   \cite{smmi}  & 1.27  & 3.37  & 3.19 & 1.79   & 1.63   & 155~ \\
SM Mi   \cite{smmi}  & 1.27  & 3.57  & 3.39 & 1.93   & 1.76   & 146~ \\
R-EDF   \cite{redf}  & 1.254 & 6.13  & 5.40 & 4.98   & 4.32   &  70~ \\ 
NR-EDF  \cite{nredf}  & 1.25  & 5.551 & 4.674 & 6.405  & 4.773  &  66~ \\   
IBM-2   \cite{ibm2}  & 1.269 & 4.68  & 3.73 & 3.70   & 3.05     &  95~ \\
QRPA Jy \cite{qrpajy}  & 1.26  & 5.26  & 3.73 & 4.00   & 2.91   &  95~ \\
QRPA CH \cite{qrpach}  & 1.25  & 5.09  &      & 1.37   & 1.55   & 118~ \\
QRPA Tu \cite{qrpatu}  & 1.27  & 5.157 & 4.642& 3.888  & 2.177  & 106~ \\
QRPA Tu \cite{qrpatu}  & 1.27  & 5.571 & 5.018& 4.373  & 2.460  &  96~ \\
\hline
\hline
\end{tabular}
\end{center}
\end{table} 
This approach is conservative since some experiments are indeed background-free or close to this regime
where sensitivities would add linearly. Each combined sensitivity  ${\cal S}$$^{comb}$ is  converted within the
respective set of NME calculations into a combined effective Majorana mass value $m_{\beta\beta}^{comb}$
(column 7 of Table~\ref{tab:nme_M}).
The lowest (66~meV) and largest (155~meV) value in column 7 of Table~\ref{tab:nme_M} define
the range of the combined effective Majorana mass shown in Table~1.


\paragraph*{Performance numbers of selected experiments}

Here we list some performance parameters
for those experiments that are used for the combined sensitivity
on $m_{\beta\beta}$.
The sensitivity  for placing a limit on
the half-life depends  on the number of expected background
events in an energy region of interest at $Q_{\beta\beta}$,
the exposure and the efficiency of reconstructing a signal
event. Parameters are often not stable
during the livetime of an experiment. We therefore list
latest values which are typically improved compared to
initial conditions.
Tighter selection cuts might improve the background at the
expense of a smaller signal efficiency and hence the
sensitivity might get worse. As a figure of merit B$_{FWHM}$ we
therefore use the  background in the energy interval of size
1$\cdot$FWHM divided
by the total efficiency $\epsilon$ (see Table~\ref{tab:comp}). 
Note: since the experiments
use different definitions for the exposure we  normalize
consistently to the active $0\nu\beta\beta$ decay isotope mass in this comparison.
\begin{table}[h]
\caption{
Characteristics of the \onbb \ experiments that have been selected for the combination
of sensitivities: energy resolution FWHM, background rate B, detection efficiency $\epsilon$,
and the figure of merit B$_{FWHM}$~=~FWHM$\times$B~/~$\epsilon$. Please note that 
background rates B are based in part on different definitions; thus a direct comparison
is not always meaningful.
}
\label{tab:comp}
\begin{center}
\begin{tabular}{lccccc}

\hline
\hline \\[-2ex]
Experiment  & isotope & FWHM & B & $\epsilon$ & B$_{FWHM}$  \\[+0.2ex]
&           & (keV) & (cts/(keV$\cdot$t$\cdot$yr)) & & (cts/(t$\cdot$yr)) \\[+0.3ex]
\hline \\[-2ex]
\gerda\ (this work) & $^{76}$Ge & 3.3 & 0.6 & 0.5 & 4 \\
\majorana \cite{majo} & $^{76}$Ge & 2.5 & 5 & 0.71 & 18 \\
Cupid-0 \cite{cupid}& $^{82}$Se & 23 & 3.6 & 0.40 & 210 \\
Cuore \cite{cuore}& $^{130}$Te & 7.4 & 14 & 0.23 & 450 \\
EXO-200 \cite{exo}& $^{136}$Xe & 71 & 1.6 & 0.66 & 170 \\
Kamland-Zen \cite{kamlzen}& $^{136}$Xe & 270 & 0.45 & 1.0 & 120 \\
\hline
\hline
\end{tabular}
\end{center}
\end{table}
														      
For \gerda\ we use the Phase~II performance numbers: 
the average resolution FWHM $\sim 3.3$~keV,
background rate is 0.6~cts/(keV$\cdot$t$\cdot$yr)
and total efficiency is $\sim 0.5$ including the active volume fraction
and isotope fraction. 

For \majorana\  updated values are available
from ref. \cite{majo}: the
resolution  FWHM = 2.5~keV,  background rate in the active
volume is 5~cts/(keV$\cdot$t$\cdot$yr) for low background condition,
the reconstruction efficiency is 0.81 and 0.88 is the enrichment fraction.

Cupid-0 reports a background rate of 3.614~cts/(keV$\cdot$t$\cdot$yr), 
the resolution is FWHM = 23 keV and the reconstruction efficiency is 0.75.

The data set 2 of Cuore has a background of 14~cts/(keV$\cdot$t$\cdot$yr)
normalized to the total mass,
a resolution of FWHM = 7.4~keV. The $^{130}$Te mass fraction is 0.28 and
the total reconstruction efficiency is 0.83.

EXO-200 quotes a background rate for the latest data set of
1.6 cts/(keV$\cdot$t$\cdot$yr), an energy resolution of FWHM = 71 keV,
an enrichment fraction of 0.8 and a reconstruction efficiency of 0.81.

For Kamland-Zen (Period 2) no background rate
is quoted. We use for the calculation the event count of 11 in a volume of
1~m radius and a 400~keV energy interval at $Q_{\beta\beta}$
as listed in \cite{kamlzen}. The total exposure is 126~kg$\cdot$yr for this
volume (Period 2 corresponds to 49\% thereof) and the resolution
is FWHM = 270~keV. The exposure includes all dead times and the
 reconstruction efficiency is 1.

Table~\ref{tab:comp} shows that it is \gerda 's combination of very good energy 
resolution and very low background rate that results in its outstanding figure of merit.

\end{document}

%% file: FIGURE1.tex
\begin{figure*}[h!]
\centering
\includegraphics[width=0.6\textwidth]{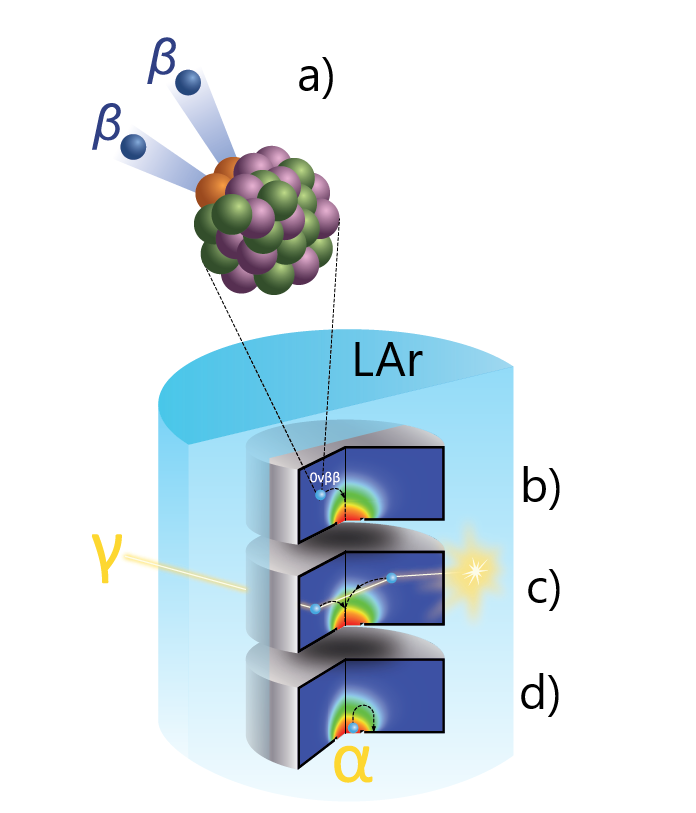}
\caption{
In \onbb \ decay two electrons ($\beta$ particles) escape the nucleus (a) and deposit energy
\qbb \ very localized in a single detector (b). 
\gerda \ searches for the decay $^{76}$Ge$\rightarrow ^{76}$Se~+~2$e^-$, \qbb \,=~2039~keV,
with high-purity Ge detectors enriched in $^{76}$Ge that are operated in 
liquid argon (LAr). 
Events with coincident LAr scintillation light or with multiple interactions in the 
Ge detector, e.g. Compton scatters (c), are classified as background events.
The special detector design with a small readout electrode enhances drift
time differences between different trajectories (black dashed lines) of the charges (holes) 
generated by the energy depositions. The color code indicates the electrical signal strength
at the respective location.
Hence single- and multi-site events can be identified efficiently by the time profile 
of their electronic signal. Similarly, $\alpha$ decays at the readout electrode (d) 
show unique signal characteristics. 
}
\end{figure*}

%% file: FIGURE2.tex
\begin{figure*}[ht!]
\centering
\includegraphics[width=0.8\textwidth]{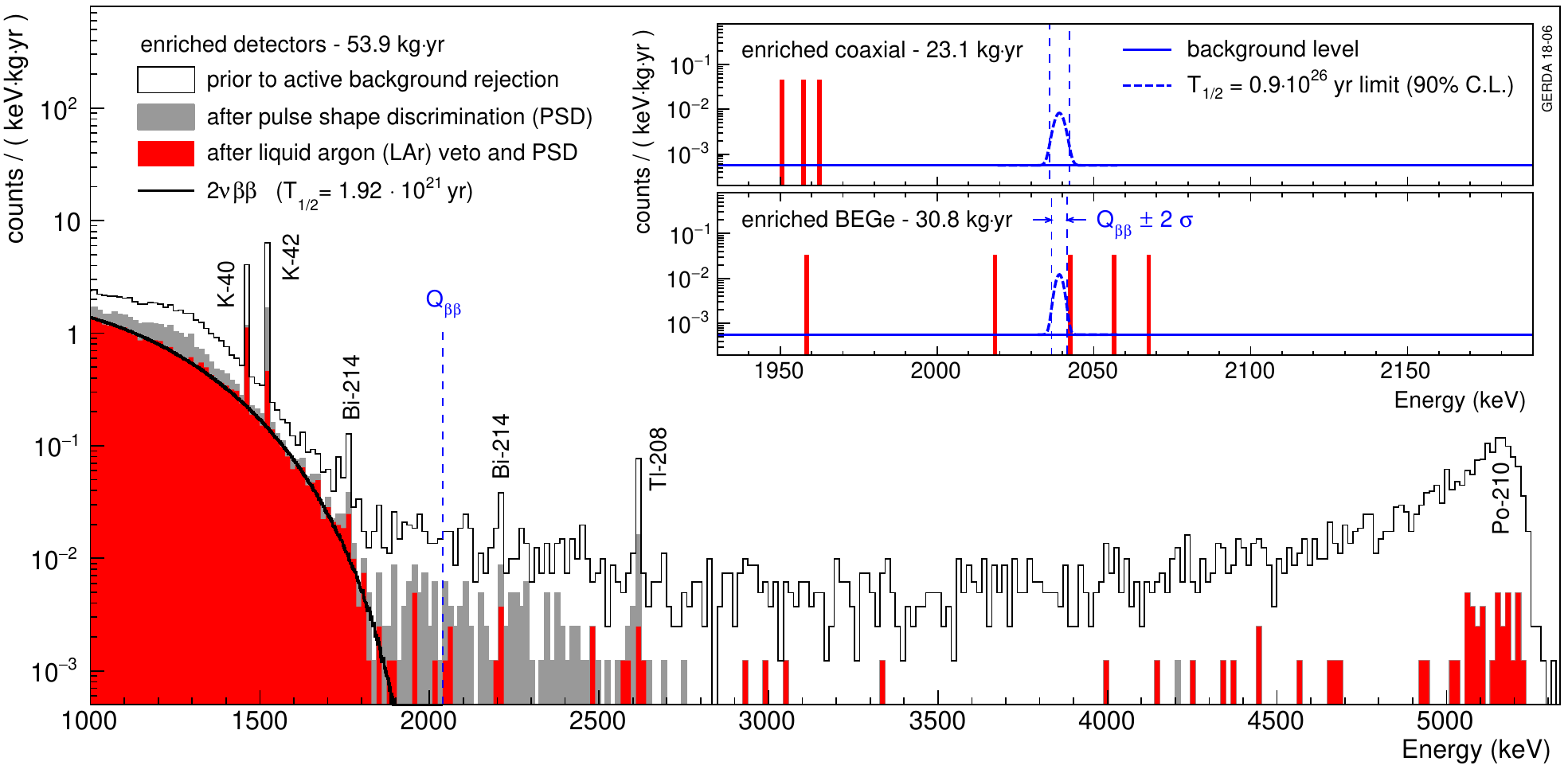}
\caption{
\gerda \ Phase II energy spectra (53.9 kg$\cdot$yr).
Enriched coaxial and BEGe data are displayed in a combined spectrum after indicated cuts.
Main contributions to the spectra are labeled. 
The insets display 
the analysis window for both coaxial and BEGe detectors separately including the background rates (solid blue lines).  
No event reconstructs within \qbb\,$\pm\,2\sigma$. 
The dashed blue curves depict the 90\% C.L. limit for a \onbb\ signal of $T^{0\nu}_{1/2} = 0.9 \cdot 10^{26}$~yr derived from the 
likelihood analysis of all \gerda \ data sets.
}
\end{figure*}

%% file: TABLE1.tex
\begin{table}[hb!]
\begin{center}
\caption{ Lower half-life limits $\cal L$(T$_{1/2}$) and sensitivities
$\cal S$(T$_{1/2}$), both at 90\% C.L., reported by recent \onbb \ decay searches with indicated deployed isotope masses 
M$_i$ and FWHM energy resolutions. 
Sensitivities $\cal S$(T$_{1/2}$) have been converted into upper limits of effective Majorana masses $m_{\beta\beta}$ 
using the nuclear matrix elements quoted in ref.~\cite{Engel:2016xgb}. 
\label{tab:limsens}
}
\vspace*{5mm}
\begin{tabular}{lrrcccr}

\hline
\hline \\[-2ex]
Experiment &         Isotope             &  M$_i$~~~& FWHM  & $\cal L$(T$_{1/2}$) & $\cal S$(T$_{1/2})$  & $m_{\beta\beta}$~~~~\\[+0.2ex]
           &                             &  (kmol)   & (keV) & (10$^{25}$\,yr) &  (10$^{25}$\,yr) & (meV)~~~\\ [+0.3ex]
\hline \\[-2ex]
GERDA (this work)          &  $^{76}$Ge~~& 0.41 &  3.3 &    9  &  11     & 104~-~228\\
Majorana \cite{majo}       & $^{76}$Ge~~&  0.34 &  2.5 &   2.7 &  4.8   & 157~-~346\\
CUPID-0 \cite{cupid}         & $^{82}$Se~~&0.063 & 23  & 0.24  &  0.23& 394~-~810\\
CUORE \cite{cuore}        & $^{130}$Te~~&1.59  & 7.4 &   1.5  &  0.7  & 162~-~757\\
EXO-200 \cite{exo}        & $^{136}$Xe~~&1.04  & 71 &  1.8   & 3.7   &~93~-~287\\[+0.2ex]
KamLAND-Zen \cite{kamlzen}& $^{136}$Xe~~&2.52  & 270 &  10.7  & 5.6 & ~76~-~234\\
\hline \\[-2ex]
Combined   &               &     &      &     &      &  66~-~155 \\[+0.2ex]
\hline
\hline
\end{tabular}
\end{center}
\end{table}

%% file: FIGURE3.tex
\begin{figure*}[b!]
\centering
	\includegraphics[width=0.9\textwidth]{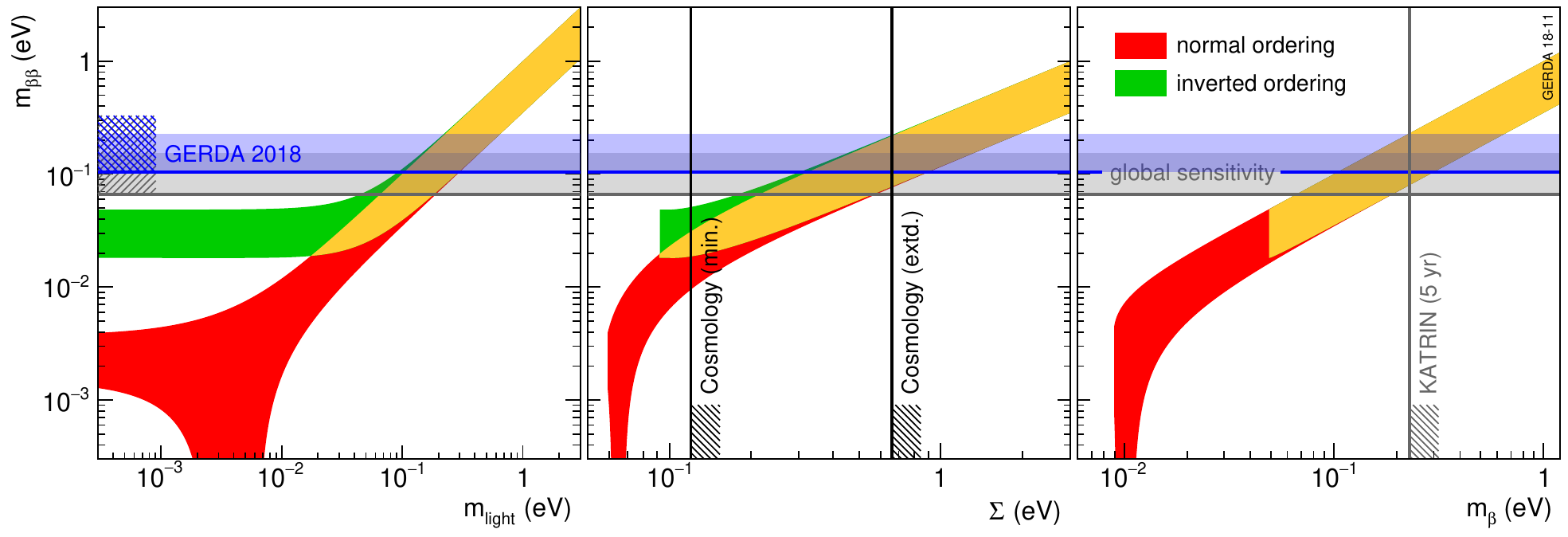}
\caption{
Constraints of the parameter space for $m_{\beta\beta}$ in the scenario of 3 light Majorana neutrinos
as function of the lightest neutrino mass $m_{light}$, the sum of neutrino masses $\Sigma$, and the effective 
neutrino mass $m_\beta$.
Contours follow from a scan of the Majorana phases with the central oscillation parameters from 
NuFIT~4.0 \cite{nufit2019}. The blue horizontal
band shows the upper limits on $m_{\beta\beta}$ obtained by \GERDA , the grey band those from combining
sensitivities of all leading experiments in the field (see Table~\ref{tab:limsens}). Vertical
lines denote $\Sigma$~=~0.12\,eV and $\Sigma$~=~0.66\,eV, a stringent limit from cosmology \cite{Aghanim:2018eyx}
and an extended model bound 
\cite{Tanabashi:2018oca},  
as well as $m_\beta$~=~0.23\,eV, the 5\,yr sensitivity of the KATRIN experiment \cite{katrin}.
Hatched areas indicate the regions of the excluded parameter space.  
 \newline \newline    
}
\end{figure*}

%% file: MSarxiv.bbl
\begin{thebibliography}{10}

\bibitem{Cowan:1992xc}
C.~L. Cowan {\it et~al.\/}, {\it Science\/} {\bf 124}, 103 (1956).

\bibitem{Fukuda:1998mi}
Y.~Fukuda {\it et~al.\/}, {\it Phys. Rev. Lett.\/} {\bf 81}, 1562 (1998).

\bibitem{Ahmad:2002jz}
Q.~R. Ahmad {\it et~al.\/}, {\it Phys. Rev. Lett.\/} {\bf 89}, 011301 (2002).

\bibitem{kzen}
K.~Eguchi {\it et~al.\/}, {\it Phys. Rev. Lett.\/} {\bf 90}, 021802 (2003).

\bibitem{Kraus:2004zw}
C.~Kraus {\it et~al.\/}, {\it Eur. Phys. J. C\/} {\bf 40}, 447 (2005).

\bibitem{Aseev:2011dq}
V.~N. Aseev {\it et~al.\/}, {\it Phys. Rev. D\/} {\bf 84}, 112003 (2011).

\bibitem{Tanabashi:2018oca}
M.~Tanabashi {\it et~al.\/}, {\it Phys. Rev. D\/} {\bf 98}, 030001 (2018).

\bibitem{Majorana:1937vz}
E.~Majorana, {\it Nuovo Cim.\/} {\bf 14}, 171 (1937).

\bibitem{moha2006}
R.N.~Mohapatra, A.Y.~Smirnov, {\it Ann. Rev. Nucl. Part. Sci.\/} {\bf 56}, 569 (2006).

\bibitem{SValle}
J.~Schechter, J.W.F.~Valle, {\it Phys. Rev. D\/} {\bf 22}, 2227 (1980).

\bibitem{Doro}
S.~Dell'Oro {\it et~al.\/}, {\it PoS(NEUTEL2017)\/} 030 (2018).

\bibitem{Mount:2010zz}
B.~J. Mount, M.~Redshaw, E.~G. Myers, {\it Phys. Rev. C\/} {\bf 81}, 032501
  (2010).

\bibitem{nature544}
M.~Agostini {\it et~al.\/}, {\it Nature\/} {\bf 544}, 47 (2017).

\bibitem{Agostini:2017hit}
M.~Agostini {\it et~al.\/}, {\it Eur. Phys. J. C\/} {\bf 78}, 388 (2018).

\bibitem{Agostini:2018tnm}
M.~Agostini {\it et~al.\/}, {\it Phys. Rev. Lett.\/} {\bf 120}, 132503 (2018).

\bibitem{gerdaPRL2013}
M.~Agostini {\it et~al.\/}, {\it Phys. Rev. Lett.\/} {\bf 111}, 122503 (2013).

\bibitem{nufit2019}
I.~Esteban {\it et~al.\/}, {\it JHEP.\/} {\bf 01}, 106 (2019).

\bibitem{Engel:2016xgb}
J.~Engel, J.~Men\'endez, {\it Rept. Prog. Phys.\/} {\bf 80}, 046301 (2017).

\bibitem{katrin}
J.~Angrik {\it et~al.\/}, KATRIN Design Report 2004 {\it FZKA Scientific Report 7090 \/}
https://www.katrin.kit.edu/publikationen/DesignReport2004-12Jan2005.pdf

\bibitem{Aghanim:2018eyx}
N.~Aghanim {\it et~al.\/}, {\it Astr. \& \ Astrophysics\/} in print, arXiv:1807.06209.

\bibitem{Abgrall:2017syy}
N.~Abgrall {\it et~al.\/}, {\it AIP Conf. Proc.\/} {\bf 1894}, 020027 (2017).

\bibitem{majo}
S.I.~Alvis {\it et~al.\/}, in print, arXiv:1902.02299.

\bibitem{cupid}
O.~Azzolini {\it et~al.\/}, {\it Phys. Rev. Lett.\/} {\bf 120 }, 232502 (2018).

\bibitem{cuore}
C.~Alduino {\it et~al.\/}, {\it Phys. Rev. Lett.\/} {\bf 120}, 132501 (2018).

\bibitem{exo}
J.B.~Albert {\it et~al.\/}, {\it Phys. Rev. Lett.\/} {\bf 120}, 072701 (2018).

\bibitem{kamlzen}
A.~Gando {\it et~al.\/}, {\it Phys. Rev. Lett.\/} {\bf 117}, 082503 (2016).


\bibitem{Ackermann:2012xja}
K.H.~Ackermann, {\it et~al.\/}, {\it Eur. Phys. J. C\/} {\bf 73}, 2330 (2013).

\bibitem{Maneschg:2008zz}
W.~Maneschg, {\it et~al.\/}, {\it Nucl. Instrum. Meth. A\/} {\bf 593}, 448 (2008).

\bibitem{Barabanov:2009zz}
I.~Barabanov, {\it et~al.\/}, {\it Nucl. Instrum. Meth. A\/} {\bf 606}, 790 (2009).

\bibitem{Freund:2016fhz}
K.~Freund, {\it et~al.\/}, {\it Eur. Phys. J. C\/} {\bf 76}, 298 (2016).

\bibitem{Gunther:1997ai}
M.~Gunther, {\it et~al.\/}, {\it Phys. Rev. D\/} {\bf 55}, 54 (1997).

\bibitem{Aalseth:2002rf}
C.E.~Aalseth, {\it et~al.\/}, {\it Phys. Rev. D\/} {\bf 65}, 092007 (2002).

\bibitem{Agostini:2013jta}
M.~Agostini, {\it et~al.\/}, {\it Eur. Phys. J. C\/} {\bf 73}, 2583 (2013).

\bibitem{Agostini:2014hra}
M.~Agostini, {\it et~al.\/}, {\it Eur. Phys. J. C\/} {\bf 75}, 39 (2015).

\bibitem{Riboldi:2015dzj}
S.~Riboldi, {\it et~al.\/}, {\it Proceedings, 4th Int. Conf. on Advancements in Nuclear Instrumentation
                                 Measurement Methods and their Applications (ANIMMA 2015)\/}, 7465549 (2015).
				 
\bibitem{Agostini:2015boa}
M.~Agostini, {\it et~al.\/}, {\it Eur. Phys. J. C\/} {\bf 75}, 506 (2015).

\bibitem{JanicskoCsathy:2010bh}
J.~Janicsk{\'o}-Cs{\'a}thy, {\it et~al.\/}, {\it Nucl. Instrum. Meth. A\/} {\bf 654}, 225 (2011).

\bibitem{Lubashevskiy:2017lmf}
A.~Lubashevskiy, {\it et~al.\/}, {\it Eur. Phys. J. C\/} {\bf 78}, 15 (2018).

\bibitem{Agostini:2011nf}
M.~Agostini, {\it et~al.\/}, {\it J. Phys. Conf. Ser.\/} {\bf 375}, 042027 (2012).

\bibitem{Agostini:2011xe}
M.~Agostini, {\it et~al.\/}, {\it JINST\/} {\bf 6}, P08013 (2011).

\bibitem{Agostini:2011mh}
M.~Agostini, {\it et~al.\/}, {\it H. Phys. Conf. Ser.\/} {\bf 368}, 012047 (2012).

\bibitem{Agostini:2015pta}
M.~Agostini, {\it et~al.\/}, {\it Eur. Phys. J. C\/} {\bf 75}, 255 (2015).

\bibitem{Baudis:2013kaa}
L.~Baudis, {\it et~al.\/}, {\it Nucl. Instrum. Meth. A\/} {\bf 729}, 557 (2013).

\bibitem{Baudis:2015sba}
L.~Baudis, {\it et~al.\/}, {\it JINST\/} {\bf 10}, P12005 (2015).

\bibitem{Agostini:2015nwa}
M.~Agostini, {\it et~al.\/}, {\it Eur. Phys. J. C\/} {\bf 75}, 416 (2015).

\bibitem{PDG}
M.~Tanabashi, {\it et~al.\/} (Particle Data Group), {\it Phys. Rev. D\/} {\bf 98}, 030001 (2018).

\bibitem{kotila}
J.~Kotila, F.~Iachello, {\it Phys. Rev. C\/} {\bf 85}, 034316 (2012).

\bibitem{stefanik}
D.~Stefanik, {\it et~al.\/}, {\it Phys. Rev. C\/} {\bf 92}, 055502 (2015).

\bibitem{smst}
J.~Men\'endez, {\it et~al.\/}, {\it Nucl. Phys. A\/} {\bf 818}, 139 (2009).

\bibitem{smmi}
M.~Horoi, A.~Neascu, {\it Phys. Rev. C\/} {\bf 93}, 024308 (2016).

\bibitem{redf}
J.M.~Yao, {\it et~al.\/}, {\it Phys. Rev. C\/} {\bf 91}, 024316 (2015).

\bibitem{nredf}
N.~L\'opez Vaquero, {\it et~al.\/}, {\it Phys. Rev. Lett.\/} {\bf 111}, 142501 (2013).

\bibitem{ibm2}
J.~Barea, {\it et~al.\/}, {\it Phys. Rev. C\/} {\bf 91}, 034304 (2015).

\bibitem{qrpajy}
J.~Hyv\"arinen, J.~Suhonen,  {\it Phys. Rev. C\/} {\bf 91}, 024613 (2015).

\bibitem{qrpach}
M.T.~Mustonen, J.~Engel, {\it Phys. Rev. C\/} {\bf 87}, 064302 (2013).

\bibitem{qrpatu}
F.~\v Simkovic, {\it et~al.\/}, {\it Phys. Rev. C\/} {\bf 87}, 045501  (2013).

\end{thebibliography}
